\documentclass[twocolumn,secnumarabic,amssymb, nobibnotes, aps, prb, superscriptaddress]{revtex4-2}
\usepackage{lipsum}
\usepackage{titlesec}
\titlespacing\section{0pt}{12pt plus 4pt minus 4pt}{1pt plus 20pt minus 2pt}
\usepackage{xcolor}
\usepackage{amsmath}
\usepackage{comment}
\usepackage{physics}
\usepackage{afterpage}
\usepackage{placeins}
\usepackage{graphicx}
\usepackage{float}
\usepackage{booktabs}
\usepackage{multirow}
\usepackage{array}
\usepackage{setspace}
\graphicspath{{Figs/}}
\usepackage{siunitx}
\usepackage{hhline}
\usepackage{xfrac}
\usepackage{float,graphicx}
\usepackage{mathtools}
\usepackage{listings}
\usepackage{amssymb}
\usepackage{titlesec}
\usepackage{amsfonts}
\usepackage[version=4]{mhchem}

\usepackage{epstopdf}
\catcode`@11
\def\seceqaa{\@addtoreset{equation}{section}
\def\theequation{A\arabic{equation}}}
\def\seceqbb{\@addtoreset{equation}{section}
\def\theequation{B\arabic{equation}}}
\def\seceqcc{\@addtoreset{equation}{section}
\def\theequation{C\arabic{equation}}}
\def\seceqdd{\@addtoreset{equation}{section}
\def\theequation{D\arabic{equation}}}
\def\seceqee{\@addtoreset{equation}{section}
\def\theequation{E\arabic{equation}}}
\def\seceqff{\@addtoreset{equation}{section}
\def\theequation{F\arabic{equation}}}
\def\seceqgg{\@addtoreset{equation}{section}
\def\theequation{G\arabic{equation}}}
\def\seceqhh{\@addtoreset{equation}{section}
\def\theequation{H\arabic{equation}}}
\catcode`@11

\begin{document}
\title{Atomic disorder and Berry phase driven anomalous Hall effect in Co$_2$FeAl Heusler compound} 

\author{Gaurav K. Shukla}
\affiliation{School of Materials Science and Technology, Indian Institute of Technology (Banaras Hindu University), Varanasi 221005, India}

\author{Ajit K. Jena}
\affiliation{Indo-Korea Science and Technology Center (IKST), Bangalore 560065, India}

\author{Nisha Shahi}
\affiliation{School of Materials Science and Technology, Indian Institute of Technology (Banaras Hindu University), Varanasi 221005, India}

\author{K. K. Dubey}
\affiliation{School of Materials Science and Technology, Indian Institute of Technology (Banaras Hindu University), Varanasi 221005, India}

\author{Indu Rajput}
\affiliation{UGC-DAE Consortium for Scientific Research, Indore, India}

\author{Sonali Baral}
\affiliation{UGC-DAE Consortium for Scientific Research, Indore, India}

\author{Kavita Yadav}
\affiliation{School of Basic Sciences,Indian Institute of Technology, Mandi, India}

\author{K. Mukherjee}
\affiliation{School of Basic Sciences,Indian Institute of Technology, Mandi, India}

\author{Archana Lakhani}
\affiliation{UGC-DAE Consortium for Scientific Research, Indore, India}

\author{Karel Carva}
\affiliation{Department of Condensed Matter Physics, Charles University, Ke Karlovu 5, CZ-12116 Praha, Czech Republic}

\author{Seung-Cheol Lee}
\affiliation{Indo-Korea Science and Technology Center (IKST), Bangalore 560065, India}

\author{Satadeep Bhattacharjee}
\affiliation{Indo-Korea Science and Technology Center (IKST), Bangalore 560065, India}

\author{Sanjay Singh}
\affiliation{School of Materials Science and Technology, Indian Institute of Technology (Banaras Hindu University), Varanasi 221005, India}

\begin{abstract}
Co$_2$-based Heusler compounds are the promising materials for the spintronics application due to their  high Curie  temperature, large spin-polarization, large magnetization density, and exotic transport properties. In the present manuscript, we report the 
anomalous Hall effect (AHE) in a polycrystalline Co$_2$FeAl Heusler compound using combined experimental  and theoretical studies.
The Rietveld analysis of high-resolution synchrotron x-ray diffraction data reveals a large degree ($\sim$ 50 \%) of antisite disorder between Fe and Al atoms. 
The analysis of anomalous transport data provides the experimental anomalous Hall conductivity (AHC) about 227 S/cm at 2\,K with an intrinsic contribution of 155 S/cm, which has nearly constant variation with temperature. The detailed scaling analysis of anomalous Hall resistivity suggests that the AHE in Co$_2$FeAl is governed by the Berry phase driven intrinsic mechanism. Our theoretical calculations reveal that the disorder present in Co$_2$FeAl compound enhances the Berry curvature induced intrinsic AHC.  
\end{abstract}
\maketitle

 Hall effect is defined as the realization of transverse electric field when a magnetic field is applied to a current-carrying conductor \cite{nagaosa2006anomalous}. Ferromagnetic materials show anomalous Hall effect (AHE) due  to the interaction between spin-orbit coupling (SOC) and magnetization. \cite{nagaosa2006anomalous,nagaosa2010anomalous,tian2009proper,yue2017towards}.
 AHE finds renewed attention in condensed matter physics due to huge application in magnetic sensors, random access memory, and spin logic devices \cite{nagaosa2010anomalous,nakatsuji2015large,hao2017anomalous}.
Two possible mechanisms have been proposed to explain the origin of AHE. An extrinsic mechanism related to the scattering events, which includes skew scattering and side jump, another one is an intrinsic mechanism related to the band structure of the material \cite{nagaosa2010anomalous,smit1955spontaneous,smit1958spontaneous}. 
The intrinsic mechanism was proposed by Karplus and Luttinger (K-L theory of intrinsic mechanism), which is connected with the role of SOC in electronic band structure of ferromagnetic material and results into the anomalous velocity of electrons perpendicular to the electric field direction \cite{karplus1954hall,sundaram1999wave,nagaosa2010anomalous}. 
Later, K-L theory was well understood in terms of Berry phase and Berry curvature \cite{xiao2010berry}. The Berry curvature is identical to a fictitious magnetic field in momentum space related to the geometrical phase of the electronic wave function\cite{manna2018heusler}. Berry curvature in momentum space introduces the transverse momentum to the electrons and gives intrinsic AHE \cite{ye1999berry,he2012berry}. 
\par
Berry curvature is highly sensitive to the electronic band structure of material and modulation in the band structure can influence the Berry curvature and hence intrinsic anomalous Hall conductivity (AHC) \cite{shen2020local}. The disorder may change the topology of the Fermi-surface or position of the Fermi level or modify the local potential environment that breaks the translational symmetry, inevitably modifying the band structure, which may reshape the AHE \cite{vidal2011influence,ernst2019anomalous,shen2020local,sakuraba2020giant,kudrnovsky2013anomalous}.
 An increased AHC has been reported in the thin film of Co$_2$MnAl$_{1-x}$Si$_x$ due to increased L2$_1$ ordering within the lattice \cite{sakuraba2020giant}. Recently, the enhancement in the AHC has been observed in Fe$_2$-based high Curie temperature Heusler compounds due to the increase in the crystal symmetry, when the system transforms from inverse Heusler to B2 type (CsCl) structure \cite{mende2021large}.
\par
Co$_2$-based full Heusler compounds got enormous attention for their half-metallic behaviour, and 100\% spin-polarization around Fermi level, which are the most prominent properties useful in spintronics devices and other memory-based  applications\cite{brown2000magnetization,galanakis2002slater,kubler2007understanding,kandpal2007calculated}. Additionally, Co$_2$-based Heusler compounds are of current interest, because of their large AHE due to the large Berry curvature linked with their band structure \cite{roy2020anomalous,reichlova2018large,li2020giant,guin2019anomalous,ernst2019anomalous}. Among Co$_2$-based Heusler compounds, Co$_2$FeAl is the most prominent candidate for the data processing and storage based applications due to its large Curie temperature($\sim$ 810\,K to 900\,K), high spin-polarization, low gilbert damping factor, and ultrafast magnetization dynamics \cite{PhysRevB.96.224425,zhang2018direct,husain2016growth,malik2021ultrafast}. As literature suggest that the Co$_2$FeAl is generally crystallizes in  B2-type \cite{husain2016growth,husain2017anomalous,wang2014structural,kogachi2006structural} disordered structure, therefore, this compound provides an opportunity for the investigation of the disorder effect on the Berry curvature and intrinsic AHC. Attempts to investigate anomalous transport in Co$_2$FeAl thin films report controversial result concerning to the origin of AHE \cite{husain2017anomalous,imort2012anomalous,wang2014structural}. 

In the present manuscript, we studied AHE in a  polycrystalline bulk Co$_2$FeAl Heusler compound. Synchrotron x-ray diffraction (SXRD) data reveals a large degree of antisite disorder between Fe and Al atoms.
 The experimental value of AHC was found to be about  227 S/cm at 2\,K and 219 S/cm at 300\,K with an intrinsic contribution of 155 S/cm. This intrinsic value of AHC is an order of magnitude larger than the theoretically predicted AHC for an ordered L2$_1$ phase of Co$_2$FeAl. Our theoretical calculations show that the antisite disorder present in Co$_2$FeAl enhances the Berry curvature induced intrinsic AHC.
\begin{figure}[t]
    \centering
    \includegraphics[width=0.5\textwidth]{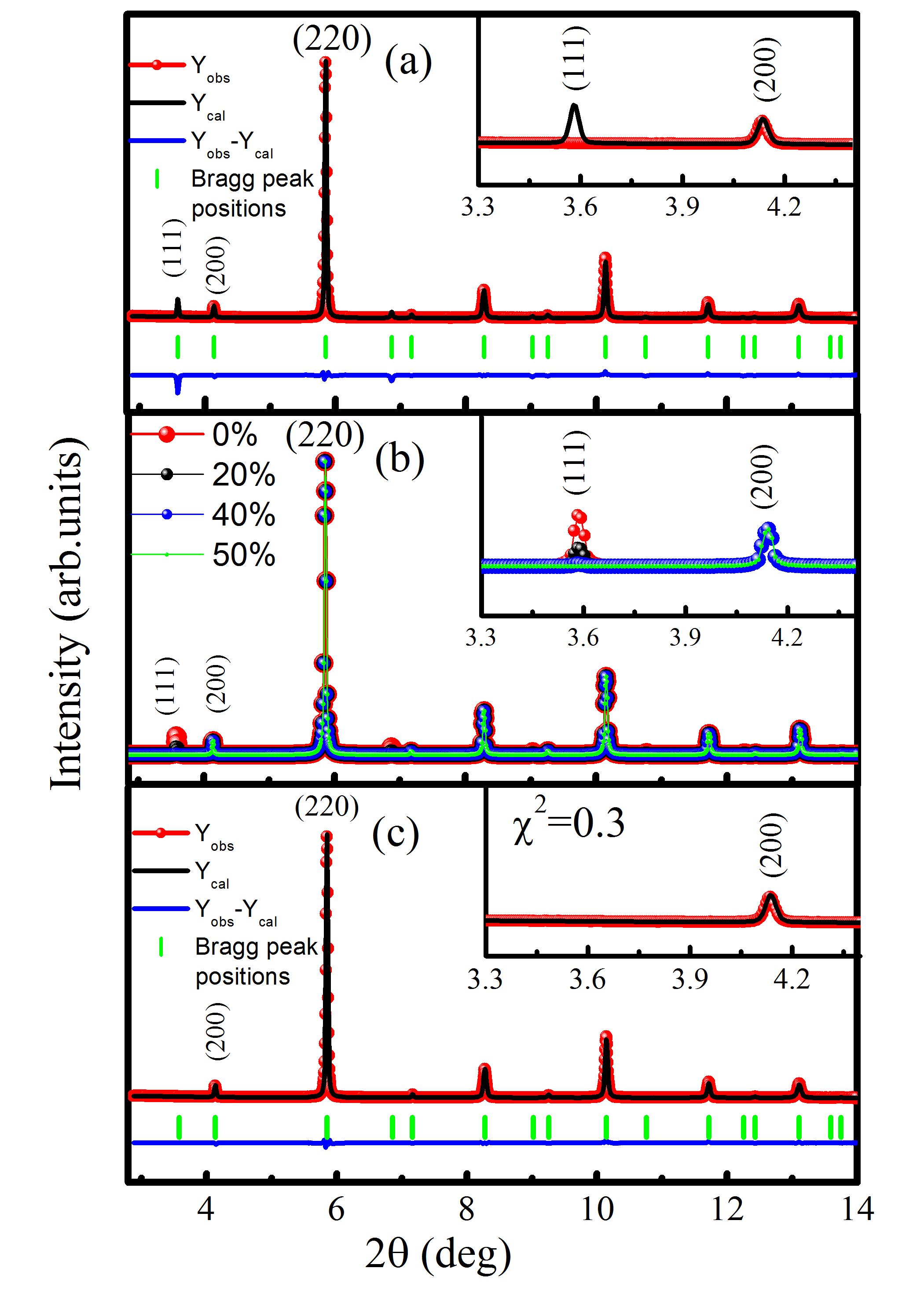}
    \caption{(a) Rietveld refinement of room temperature (RT) synchrotron x-ray diffraction (SXRD) pattern of Co$_2$FeAl considering ordered L2$_1$ structure. (b) Simulated XRD patterns considering  Fe and Al antisite disorder with indicated percentage. (c) Rietveld refinement  of RT SXRD pattern of Co$_2$FeAl with 50\% Fe and Al antisite disorder. Insets of figures show an enlarged view around (111) and (200) superlattice reflections.}
    \label{Fig1}
\end{figure}
\begin{figure}[htb]
    \centering
    \includegraphics[width=0.5\textwidth]{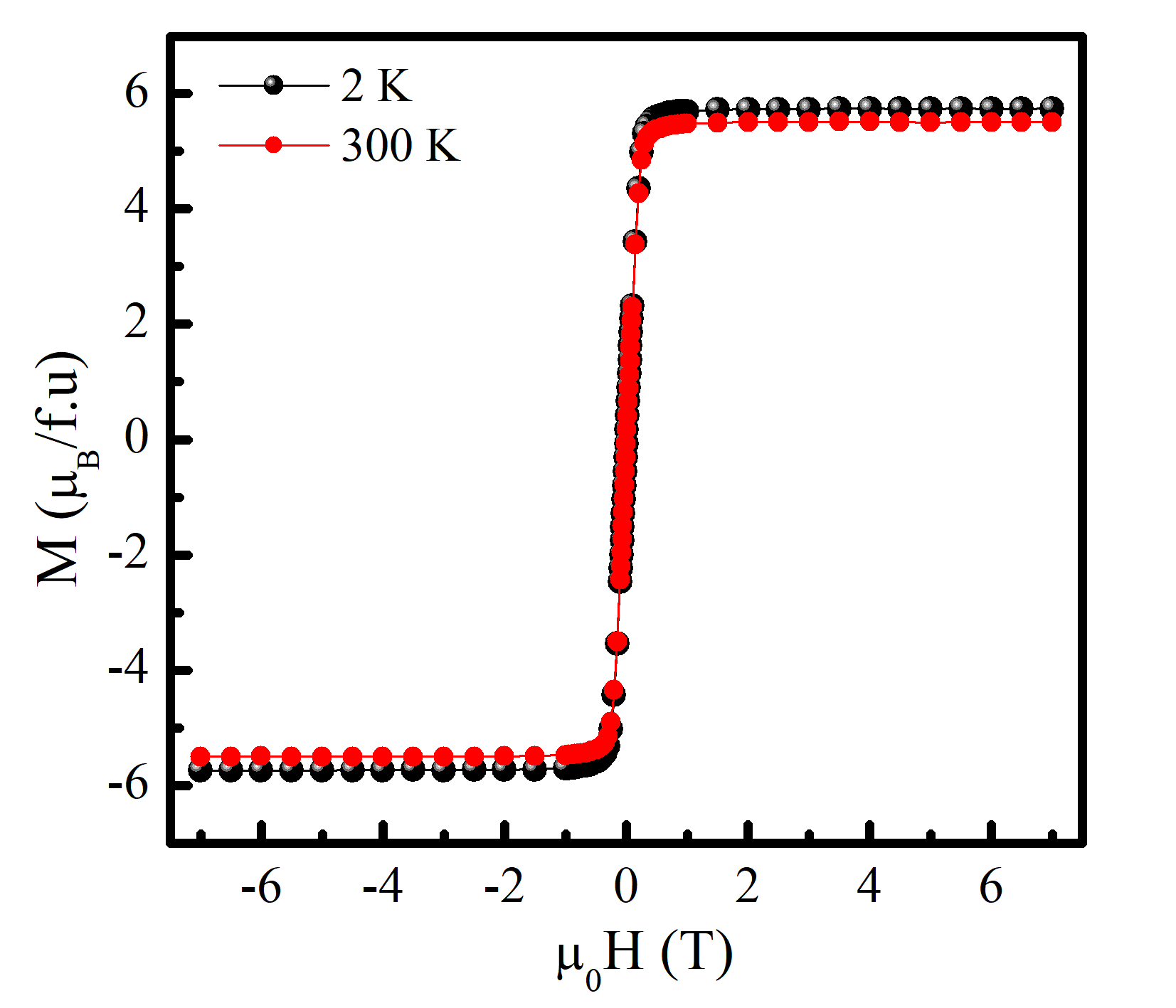}
    \caption{ Magnetic isotherms (M-H curves) recorded  at 2\,K and 300\,K.}
    \label{Fig2}
\end{figure}
\begin{figure*}[htb]
    \centering
    \includegraphics[width=1\textwidth]{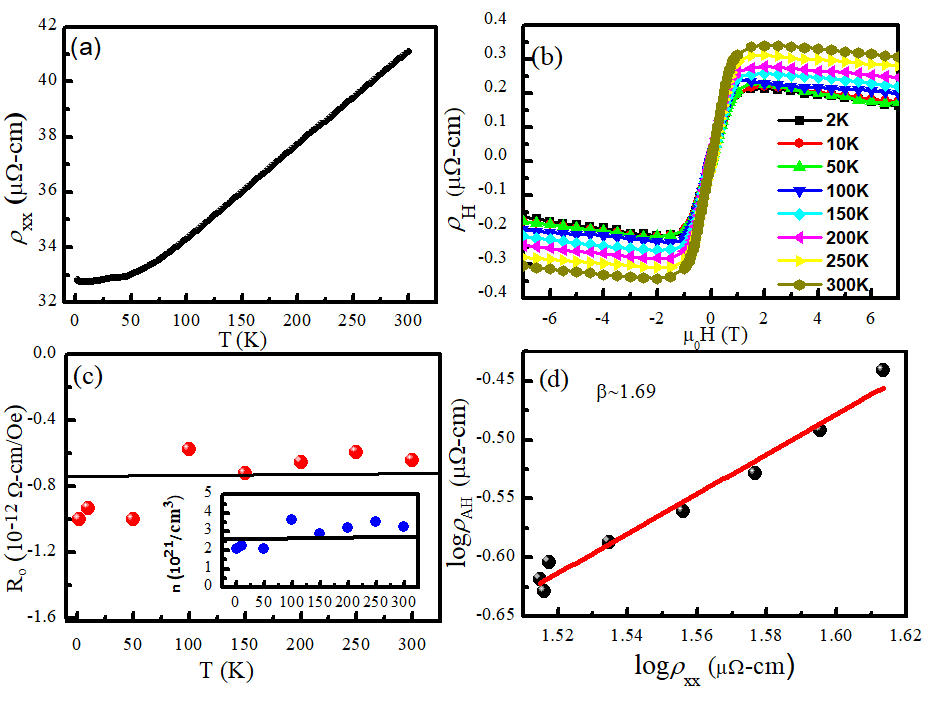}
    \caption{(a) Temperature dependent longitudinal resistivity \si{\rho}\textsubscript{\tiny{xx}}. (b) Field dependent Hall resistivity \si{\rho}\textsubscript{\tiny{H}} at different temperatures. (c) Temperature dependent normal Hall coefficient R$_0$. Inset shows temperature variation of carrier concentration n. (d) Experimental data (black dots) plotted between log\si{\rho}\textsubscript{\tiny{AH}} and log\si{\rho}\textsubscript{\tiny{xx}} and the fitted curve is shown in red color.}
    \label{Fig3}
\end{figure*}

A polycrystalline Co$_2$FeAl Heusler compound was synthesized using conventional arc melting technique using stoichiometric amount of its high pure constituent elements. The compound was melted four times to ensure the chemical homogeneity. A very small wight loss of 0.32\% was noted after melting. Further, the ingot was sealed in quartz ampoule under Ar atmosphere and then annealed at 800$^{\circ}$C for 12 hours for the better homogeneity. The energy dispersive x-ray (EDX) analysis reveals the composition ratio of 2:1:1  within the standard deviation (3 to 5 \%) of EDX measurement.
A small piece was cut from the annealed ingot and crushed into powder for SXRD measurement. The SXRD measurement was performed at PETRA-III, DESY for structural analysis using wavelength of $\SI{0.207}{\angstrom}$. Magnetic field-dependent magnetization measurements were carried out using the Magnetic Property Measurement System (MPMS) from Quantum Design, U.S.A. A small polished rectangular piece was used for four-probe and five-probe magneto-transport measurements to obtain the longitudinal resistivity (\si{\rho}\textsubscript{\tiny{xx}}) and the Hall resistivity (\si{\rho}\textsubscript{\tiny{H}}), respectively. To obtain the actual \si{\rho}\textsubscript{\tiny{H}}, raw Hall resistivity data (\si{\rho}\textsubscript{\tiny{H}}$^{raw}$) was anti-symmetrized by averaging the difference of \si{\rho}\textsubscript{\tiny{H}}$^{raw}$ at the positive and negative magnetic fields.

 Electronic structure calculations were carried out using pseudo-potential based density-functional theory and plane-wave basis sets as implemented in Quantum ESPRESSO (QE) \cite{giannozzi2009quantum}, whereas the exchange-correlation potential is approximated through PBE-GGA functional \cite{perdew1996generalized}. Optimized norm-conserving Vanderbilt pseudo-potentials \cite{hamann2013optimized} are used in the calculations and the kinetic energy cutoff for the plane-wave is taken as 80 $Ry$. The electronic integration over the Brillouin zone is approximated by the Gaussian smearing of 0.01 $Ry$ both for the self-consistent (SC) and non-self-consistent (NSC) calculations. The threshold for the SC energy calculations is taken as $10^{-8} Ry$. The projections of Bloch wave functions are made into maximally localized Wannier functions. Wannier90 tool (implemented within QE) has been used to compute the Wannier interpolated bands and AHC \cite{giannozzi2009quantum,marzari1997maximally,souza2001maximally}. SOC is introduced in all the calculations. The Monkhorst-Pack \textbf{k}-grid of $8\times8\times8$ are considered in the SC, NSC and Wannier90 calculations. The transition metal-$d$ and $Al$-$p$ orbitals are used as the projections for the Wannier90 calculations. The AHC calculation is carried out with a dense \textbf{k}-grid of $75\times75\times75$. Further, through the  adaptive refinement technique a fine mesh of $5\times5\times5$ is added around the points wherever the mode of the Berry curvature ($\abs{\Omega(\textbf{k})}$) exceeds 100 bohr$^2$. The calculations are carried out using experimental lattice parameter.


\begin{figure}[htb]
    \centering
    \includegraphics[width=0.5\textwidth]{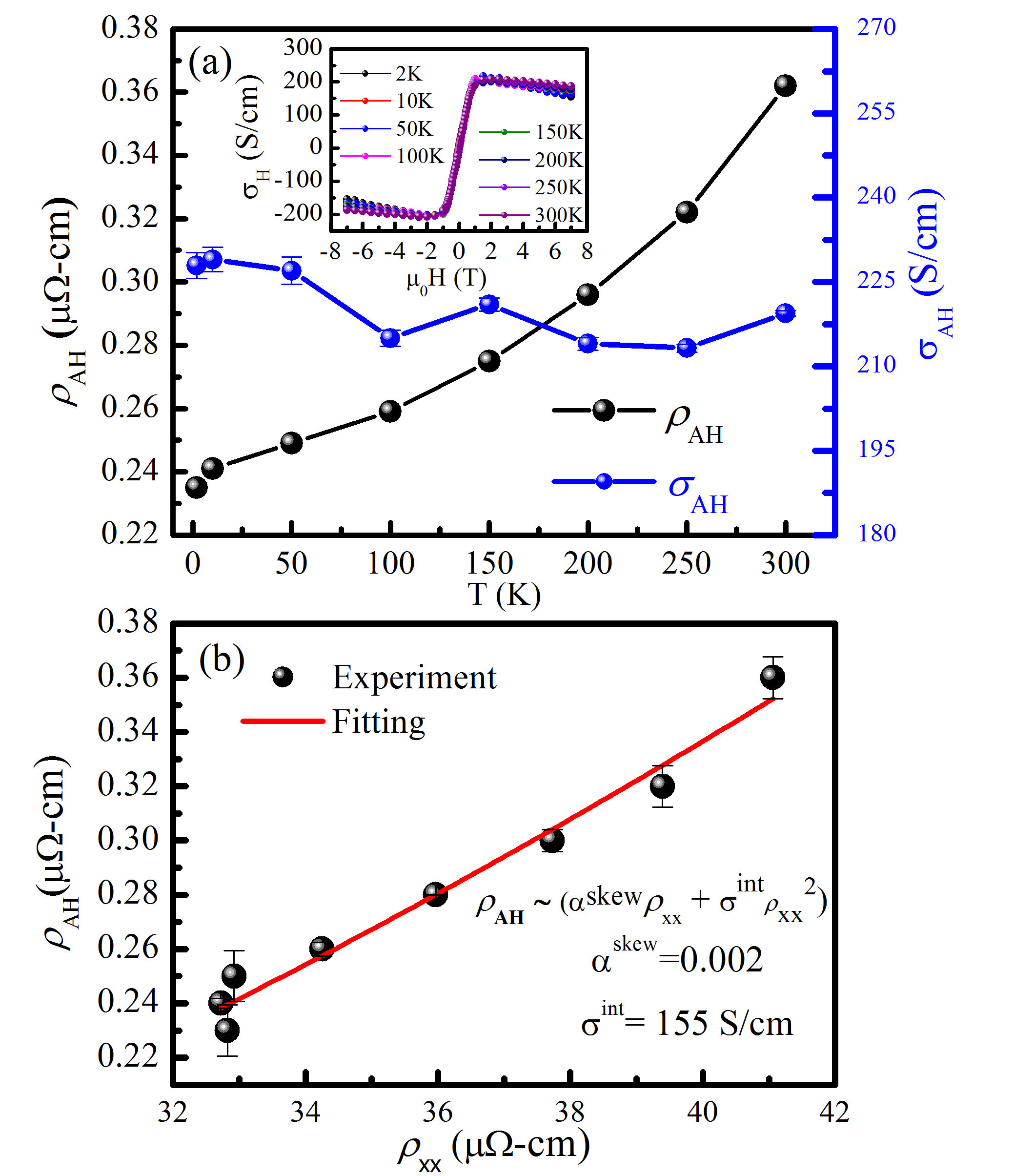}
    \caption{(a) Anomalous Hall resistivity \si{\rho}\textsubscript{\tiny{AH}} and anomalous Hall conductivity \si{\sigma}\textsubscript{\tiny{AH}}
 as a function of temperature. Inset shows field dependent  Hall conductivity isotherms. (b) Experimental
 data (black dots) plotted as \si{\rho}\textsubscript{\tiny{AH}} versus \si{\rho}\textsubscript{\tiny{xx}}. The fitted curve is shown in red color.}
    \label{Fig4}
\end{figure}
The SXRD pattern of Co$_2$FeAl compound was collected at room temperature for detail structural study. In the first step, the Rietveld refinement of SXRD pattern was carried out using the L2$_1$ ordered cubic structure with space group Fm$\bar{3}$m. 
 For the refinement, all the atoms were considered at special positions i.e. Co at 8c (0.25,0.25,0.25), Fe at 4b (0.5, 0.5, 0.5) and Al at 4a (0, 0, 0) Wyckoff positions, respectively. The result of refinements is shown in FIG.\ref{Fig1}(a).
We noticed the presence of (111) superlattice reflection in the calculated x-ray diffraction (XRD) pattern (black lines of FIG. \ref{Fig1}(a)), while this reflection is completely absent in the observed XRD pattern (red dots in FIG. \ref{Fig1}(a)), which indicates the presence of antisite disorder in the Co$_2$FeAl compound. Recently the mixed L2$_1$ and B2 phase was observed in Co$_2$FeAl ultrathin film \cite{adom}.
We would like to mention here that attempt to anneal the Co$_2$FeAl at different temperature could not show the different XRD pattern as compared to the observed in FIG. \ref{Fig1}(a).
It is important to remark here that atomic disorder is a common phenomenon in Heusler compounds \cite{vidal2011influence,hazra2018effect,sakuraba2020giant}.  The available literature also suggests that the most stable structure of Co$_2$FeAl is the B2 type structure i.e. there is antisite disorder between Fe and Al atoms \cite{husain2016growth,husain2017anomalous,wang2014structural,kogachi2006structural}. So, in the next step we simulated the XRD pattern of Co$_2$FeAl considering Fe-Al antisite disorder in such a way that total number of Fe and Al atoms remain same. For the XRD simulation, we used PowderCell software \cite{kraus1996powder}. It is clear from FIG.\ref{Fig1}(b), the intensity of (111) peak decreases with increase in amount of disorder and vanishes completely about 50\%  antisite disorder between Fe-Al atoms.
Therefore, finally we performed the Rietveld refinement of the SXRD data assuming 50\% antisite disorder between Fe-Al atoms, which could fit the  Bragg peaks very well (FIG.\ref{Fig1}(c)) and confirms the phase purity (cubic) as well as large antisite disorder (B2 type) in the sample. Moreover, the presence of (200) Bragg peak primarily indicates the formation of ordered Co-sublattice and also precludes the possibility of A2 disorder (atomic disorder among all sites) in the sample. The intensity ratio of superlattice reflection (200) and the fundamental reflection (220) ( i.e. $\frac{I_{200}}{I_{220}}$ ) was found to be 0.048 and 0.046 from the experimental SXRD pattern and simulated XRD pattern, respectively are nearly  same and further confirms the ordered Co-sublattice in the present compound \cite{husain2017anomalous}. The refined unit cell parameter was found to be $\SI{5.73}{\angstrom}$, which is in well agreement with the literature \cite{ortiz2011static,wurmehl2006electronic}. The magnetic moment obtained from the magnetic isotherms (FIG.\ref{Fig2}) is about 5.74 $\mu_B$/f.u and 5.50 $\mu_B$/f.u at 2\,K and 300\,K, respectively, which is close to the value reported in the literature \cite{kogachi2006structural,wurmehl2006electronic,jain2014electronic,husain2016growth,zhang2018atomic,ahmad2019size}. The variation of \si{\rho}\textsubscript{\tiny{xx}} as a function of temperature (FIG.\ref{Fig3}(a)) depicts that \si{\rho}\textsubscript{\tiny{xx}} increases with increasing  temperature, indicating the metallic character of the compound. The residual resistance ratio (RRR = $\frac{\si{\rho}\textsubscript{\tiny{xx}}(300 \,K)}{\si{\rho}\textsubscript{\tiny{xx}} (2\,K)})$ about 1.25 is similar to the reported value for other Co$_2$-based Heusler compounds \cite{roy2020anomalous,hirohata2006heusler,markou2019thickness}.

We carried out detailed magneto-transport measurements in a wide temperature range of 2\,K-300\,K to study the AHE in the Co$_2$FeAl. The Hall resitivity (\si{\rho}\textsubscript{\tiny{H}}) can be given by the equation, \si{\rho}\textsubscript{\tiny{H}} = R$_0$H + R$_s$M, where R$_0$, R$_s$ are the normal and anomalous Hall coefficients respectively.
\begin{figure}[t]
\includegraphics[width=0.5\textwidth ]{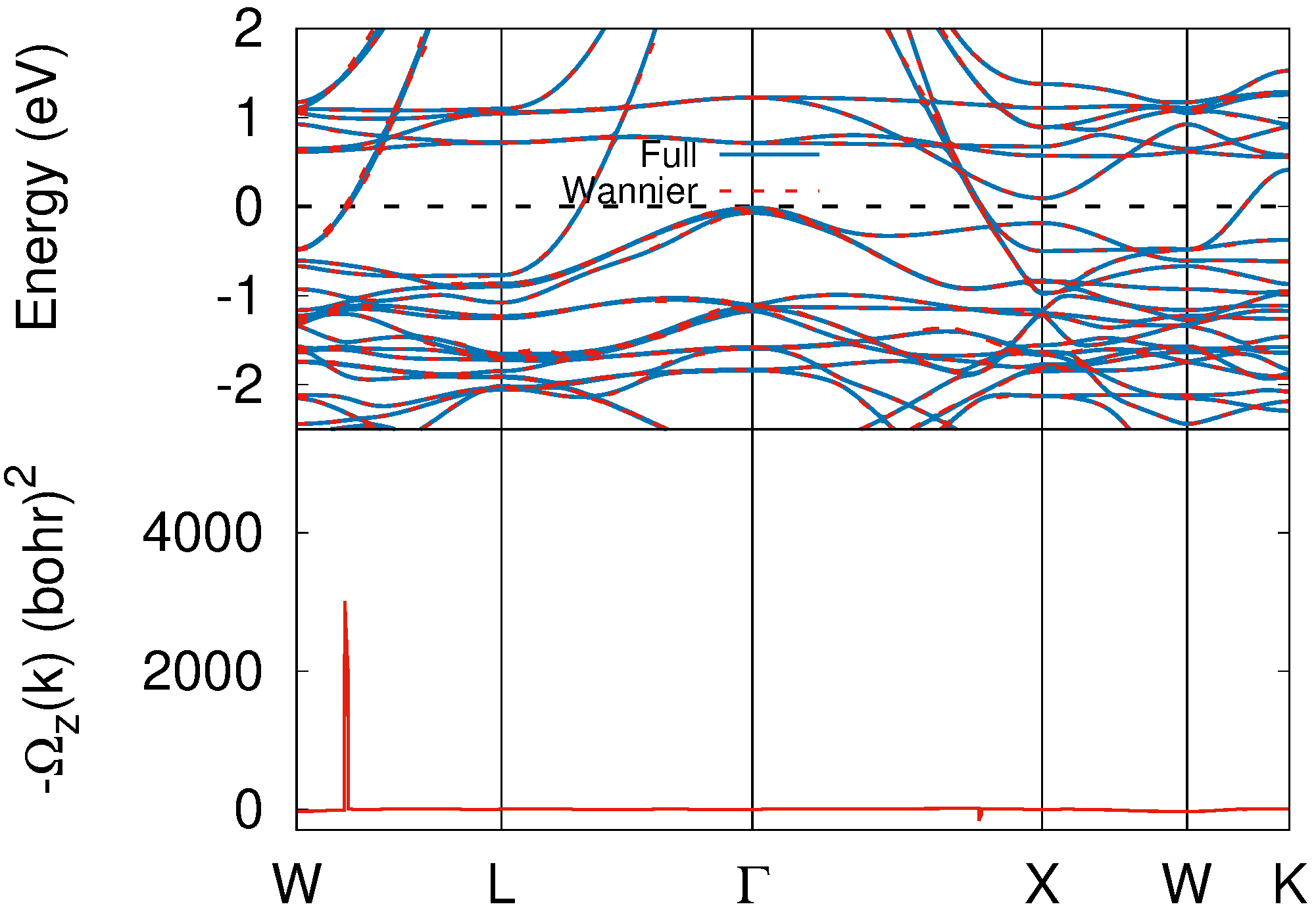}
\caption{Top: Comparison of Wannier interpolated band structure (red) with the full electronic band structure (blue) of Co$_2$FeAl. The Fermi energy is set to 0 eV. Bottom: Calculated Berry curvatures along the high-symmetry path.}
\label{fig5}
\end{figure}
\si{\rho}\textsubscript{\tiny{H}} versus H were measured at different temperatures upto field of 7 Tesla (T). From Fig.\ref{Fig3}(b), it is evident that \si{\rho}\textsubscript{\tiny{H}} steeply increases with field upto 1 T, which can be observed due to AHE. However, under the application of higher fields ($>$1\,T), a negative slope is noted due to ordinary Hall effect.
The normal Hall coefficient(R$_0$) was calculated from the slope of the high field \si{\rho}\textsubscript{\tiny{H}} curve. Fig.\ref{Fig3}(c) shows temperature variation of R$_0$. The negative value of R$_0$ indicates the electrons are the majority charge carriers.
Inset of FIG.\ref{Fig3}(c) shows the magnitude of carrier concentration (n) at different temperatures, calculated using the relation, R$_o$=$\frac{1}{ne}$ and n was found to be around 3$\times$10$^{21}$ at 300\,K and variation of n with temperature is little scattered.
 The anomalous Hall resistivity (\si{\rho}\textsubscript{\tiny{AH}}) was calculated by extrapolating the high field \si{\rho}\textsubscript{\tiny{H}} curve on y axis at zero field. 

In order to elucidate the mechanism giving rise to AHE, we have plotted \si{\rho}\textsubscript{\tiny{AH}} versus \si{\rho}\textsubscript{\tiny{xx}} on a double logarithmic scale and fitting was employed to determine the exponent $\beta$ using the relation \si{\rho}\textsubscript{\tiny{AH}} $\propto$ \si{\rho}\textsubscript{\tiny{xx}}$^{\beta}$ \cite{roy2020anomalous} as shown in FIG.\ref{Fig3}(d). If $\beta$=1, the origin of AHE is assigned to the skew scattering and if $\beta$ =2, the origin of AHE is due to intrinsic and side jump mechanisms \cite{roy2020anomalous,nagaosa2010anomalous}.  We found the exponent $\beta=1.69$, which indicates that the AHE in Co$_2$FeAl is dominated by the intrinsic and side jump mechanisms. The contribution of side jump in AHC can be estimated using an expression (e$^2$/(ha)({$\epsilon$}\textsubscript{so}/E\textsubscript{F}), where {$\epsilon$}\textsubscript{so} is the spin–orbit interaction and E\textsubscript{F} is Fermi energy \cite{nozieres1973simple,PhysRevLett.97.126602}. The terms e, h and a are the electronic charge, Planck constant and lattice parameter, respectively. For the most of the ferromagnetic metals $\epsilon$\textsubscript{so}/E\textsubscript{F} is order of $10^{-2}$, and hence very small contribution of AHC is expected due to side jump in comparison to the intrinsic part of AHC. However, it is not possible to decouple the intrinsic and side jump mechanism practically because both have similar dependencies on \si{\rho}\textsubscript{\tiny{xx}}.

We have calculated the Hall conductivity using tensor conversion \si{\sigma}\textsubscript{\tiny{H}} = $\frac{\si{\rho}\textsubscript{\tiny{H}}}{(\si{\rho}\textsubscript{\tiny{H}}^2+\si{\rho}\textsubscript{\tiny{xx}}^2)}$ \cite{manna2018colossal,hazra2018effect} as shown in inset of FIG. \ref{Fig4}(a). The AHC is calculated by averaging of extrapolated values of the high field Hall conductivity curve at zero field of the positive and negative field directions. Temperature dependent \si{\rho}\textsubscript{\tiny{AH}} (black dots) and AHC (blue dots) are shown in FIG.\ref{Fig4}(a).  The value of AHC is found to be about 227 S/cm at 2\,K and does not show appreciable change at 300\,K (219 S/cm). The variation of AHC is nearly temperature-independent, suggests that the origin of AHE is intrinsic \cite{wang2018large,liu2019magnetic}. 

To separate the extrinsic and intrinsic part of AHE, we have plotted \si{\rho}\textsubscript{\tiny{AH}} versus \si{\rho}\textsubscript{\tiny{xx}} and fitted (FIG.\ref{Fig4}(b)) according to well establish equation for AHE, $\si{\rho}\textsubscript{\tiny{AH}} = {\alpha^{skew}}\si{\rho}\textsubscript{\tiny{xx}}+\si{\sigma^{int}}\si{\rho}\textsubscript{\tiny{xx}}^2$.
where $\alpha^{skew}$ and \si{\sigma^{int}} correspond to skew scattering parameter and intrinsic AHC respectively. \si{\sigma^{int}} was estimated $\sim$ 155 S/cm, which is about 70\% of total AHC at 2\,K. Thus, in the present system intrinsic Berry phase driven K-L contribution dominates along with finite skew scattering \cite{nagaosa2010anomalous,liu2019magnetic,wang2018large,chen2019pressure}. 
 
After obtaining the experimental value of AHC, we have theoretically calculated  AHC for 
Co$_2$FeAl by setting the magnetization direction along [001]. For Co$_2$-based Heusler alloys the ground state energy in other magnetization direction like [110] was found close to the [001] direction and the band structure was also found quite similar in both directions, therefore the average picture of AHC is expected close to [001] direction \cite{CTS,Weyl,li2020giant,ernst2019anomalous}. The intrinsic AHC is proportional to the Brillouin zone (BZ) summation of the Berry curvature over all occupied states. \cite{kubler2012berry}

\begin{equation}
\sigma^{\alpha\beta}= \frac{e^2}{\hbar} \frac{1}{N}\sum_{\textbf{k}\in(BZ)} \Omega_\gamma(\textbf{k}) f(\textbf{k}),
\label{eq:ahc}
\end{equation}

where the indices $\alpha$, $\beta$, and $\gamma$ are the Cartesian coordinates. $f(\textbf{k})$ stands for the Fermi distribution function, $\Omega_\gamma(\textbf{k})$ denotes the $\gamma$ component of the Berry curvature for the wave vector $\textbf{k}$ and $N$ is the number of electrons in the crystal. Further, the Berry curvature is related to the Berry connection ($A_n(\textbf{k})$) as 

\begin{equation}
\Omega_n(\textbf{k})= \nabla_\textbf{k} \times A_n(\textbf{k}),
\label{eq:curvature}
\end{equation}

where "$n$" is the band index and $A_n(\textit{k})$ in terms of cell-periodic Bloch states $\ket{u_{n\textbf{k}}} = e^{-i\textit{k.r}}\ket{\psi_{n\textbf{k}}}$ is defined as $ A_n(\textbf{k})$= 
$\bra{u_{n\textbf{k}}}i\grad_k\ket{u_{n\textbf{k}}}$ \cite{pizzi2020wannier90}.
In the first step of the AHC calculation, we considered the ordered L2$_1$ structure of Co$_2$FeAl i.e without any disorder.
As discussed earlier the intrinsic AHC of a system is strongly connected to its electronic band structure. In FIG. \ref{fig5} (Top), we have compared the full electronic band structure of L2$_1$ ordered Co$_2$FeAl with the Wannier interpolated one. The better interpolation suggests that it will provide Wannier90 related properties accurately. 
The Wannier interpolation is a potential tool to calculate the momentum space integrals of rapidly varying functions \cite{tsirkin2021high}. Such integrals are involved in calculating the properties such as anomalous Hall conductivity, spin Hall conductivity, orbital magnetization and optical properties \cite{MOSTOFI2008685}. The most popular technique to construct the Wannier functions is maximally localized method \cite{PhysRevB.56.12847} which is implemented in the Wannier90 code \cite{MOSTOFI2008685}. The Wannier functions are generated using the unitary transformation of Bloch wave function, so there is no loss of information during the generation. The main advantage of Wannier interpolation over other approaches is that it allows for the most precise interpolation of band energies and matrix elements compared to other methods such as the tight binding approach because there is no limitation in terms of the size of the basis set \cite{wannierimp}.

\begin{figure}[t]
\includegraphics[width=0.5\textwidth ]{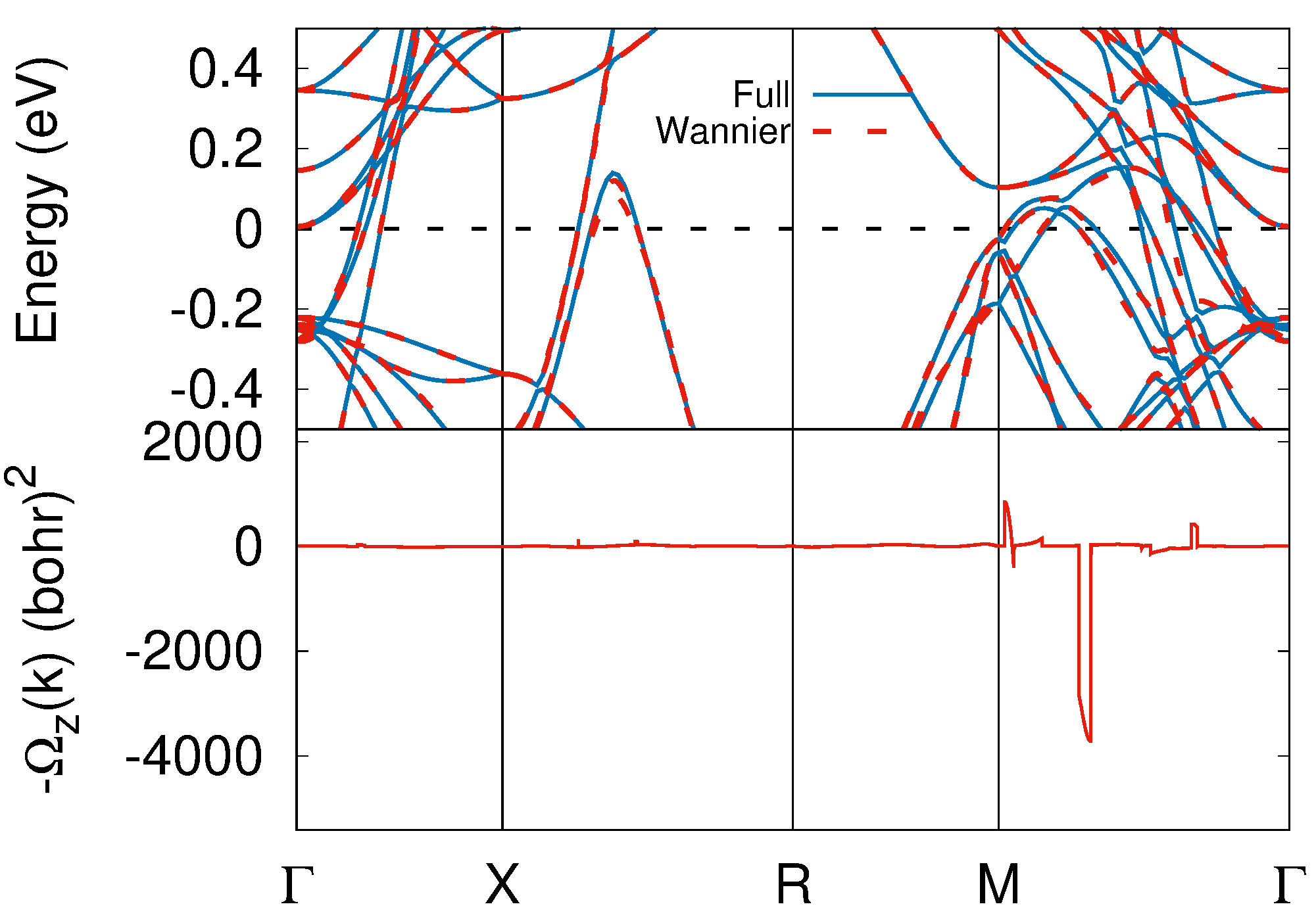}
\caption{Top: Comparison of Wannier interpolated band structure (red) with the full electronic band structure (blue) of disordered Co$_2$FeAl. The Fermi energy is set to 0 eV. Bottom: Calculated Berry curvatures along the high-symmetry path.}
\label{fig6}
\end{figure}

By this method, we found the theoretical value of AHC ($\sigma^{xy}$) about $\sim$ 42 S/cm, which is in well agreement with literature \cite{huang2015anomalous}. Thus the theoretical AHC considering ordered L2$_1$  structure is an order smaller than the experimental intrinsic AHC.
 Therefore, in the next step we incorporated 50 $\%$ anti-site disorder between the Al and Fe sites  (as observed from SXRD analysis) in order to compute AHC for the disordered Co$_2$FeAl. 
 In FIG. \ref{fig6} (Top), we have plotted the full electronic band structure of disordered Co$_2$FeAl with the Wannier interpolated band structure. The Berry curvature along high symmetry path of disordered structure (space group Pm$\bar{3}$m) is depicted in FIG. \ref{fig6} (Bottom).
 The intrinsic AHC ($\sigma^{xy}$) for disordered Co$_2$FeAl calculated from the integration of Berry curvature turned out to be $\sim$ 63 S/cm, which is interestingly larger than the ordered L2$_1$ structure. Thus, our theoretical calculations 
 suggest that the disorder can modify the Berry curvature and result into an increased value of intrinsic AHC. Recently, it has been suggested in literature that the presence of B2 disorder lowers the value of AHC in comparison to ordered L2$_1$ structure\cite{sakuraba2020giant,mende2021large}. 
Therefore, our combined experimental and theoretical results suggest that there is no straightforward rule that connects the Berry curvature to the disorder, rather it depends on the disorder induced change in electronic structure, which is different for different Heusler compound as they have different number of electrons. If the effect of disorder is such a way that it brings the band crossings or avoided band crossing very close to the Fermi energy then the value of Berry curvature will be large. We would also like to mention here that the experimentally found intrinsic AHC (155 S/cm) is larger than the theoretically predicted intrinsic AHC similar to the other metallic compounds \cite{ernst2019anomalous,zhu2020exceptionally,huang2015anomalous,wang2018large}.  
Hence, our results provide a platform for the systematic investigate of AHE in the disordered Heusler compounds and related materials.
 
In conclusion, we investigated the anomalous transport properties of polycrystalline Co$_2$FeAl Heusler compound by experiment and theoretical calculations. SXRD data reveals a large degree of Fe-Al antisite disorder. Experimental values of AHC were found to be 227 S/cm at 2\,K and 219 S/cm at 300\,K with an intrinsic AHC of 155 S/cm. Our experimental analysis show that the AHE in Co$_2$FeAl is dominated by intrinsic Berry phase mechanism. 
Our theoretical calculations suggest that the enhanced Berry curvature induced intrinsic AHC is linked with the antisite disorder present in Co$_2$FeAl Heusler compound.

We gratefully acknowledge UGC-DAE CSR for experimental support. SS thanks Science and Engineering Research Board of India for financial support through the award of Ramanujan Fellowship (grant no: SB/S2/RJN-015/2017) and Early Career Research Award (grant no: ECR/2017/003186) and UGC-DAE CSR, Indore for financial support through “CRS” Scheme. GKS thanks to DST INSPIRE scheme for financial support.Portions of this research were conducted at the light source PETRA III of DESY, a member of the Helmholtz Association. Financial support from the Department of Science and Technology, Government of
India within the framework of the India@DESY is gratefully
acknowledged. We would like to thank the beamline scientist
Dr. Martin Etter for his help in setting up the experiments.

\begin{thebibliography}{61}%
\makeatletter
\providecommand \@ifxundefined [1]{%
 \@ifx{#1\undefined}
}%
\providecommand \@ifnum [1]{%
 \ifnum #1\expandafter \@firstoftwo
 \else \expandafter \@secondoftwo
 \fi
}%
\providecommand \@ifx [1]{%
 \ifx #1\expandafter \@firstoftwo
 \else \expandafter \@secondoftwo
 \fi
}%
\providecommand \natexlab [1]{#1}%
\providecommand \enquote  [1]{``#1''}%
\providecommand \bibnamefont  [1]{#1}%
\providecommand \bibfnamefont [1]{#1}%
\providecommand \citenamefont [1]{#1}%
\providecommand \href@noop [0]{\@secondoftwo}%
\providecommand \href [0]{\begingroup \@sanitize@url \@href}%
\providecommand \@href[1]{\@@startlink{#1}\@@href}%
\providecommand \@@href[1]{\endgroup#1\@@endlink}%
\providecommand \@sanitize@url [0]{\catcode `\\12\catcode `\$12\catcode
  `\&12\catcode `\#12\catcode `\^12\catcode `\_12\catcode `\%12\relax}%
\providecommand \@@startlink[1]{}%
\providecommand \@@endlink[0]{}%
\providecommand \url  [0]{\begingroup\@sanitize@url \@url }%
\providecommand \@url [1]{\endgroup\@href {#1}{\urlprefix }}%
\providecommand \urlprefix  [0]{URL }%
\providecommand \Eprint [0]{\href }%
\providecommand \doibase [0]{https://doi.org/}%
\providecommand \selectlanguage [0]{\@gobble}%
\providecommand \bibinfo  [0]{\@secondoftwo}%
\providecommand \bibfield  [0]{\@secondoftwo}%
\providecommand \translation [1]{[#1]}%
\providecommand \BibitemOpen [0]{}%
\providecommand \bibitemStop [0]{}%
\providecommand \bibitemNoStop [0]{.\EOS\space}%
\providecommand \EOS [0]{\spacefactor3000\relax}%
\providecommand \BibitemShut  [1]{\csname bibitem#1\endcsname}%
\let\auto@bib@innerbib\@empty
\bibitem [{\citenamefont {Nagaosa}(2006)}]{nagaosa2006anomalous}%
  \BibitemOpen
  \bibfield  {author} {\bibinfo {author} {\bibfnamefont {N.}~\bibnamefont
  {Nagaosa}},\ }\bibfield  {title} {\bibinfo {title} {Anomalous \ce{Hall}
  effect—\ce{A} new perspective},\ }\href@noop {} {\bibfield  {journal}
  {\bibinfo  {journal} {J. Phys. Soc. Jpn.}\ }\textbf
  {\bibinfo {volume} {75}},\ \bibinfo {pages} {042001} (\bibinfo {year}
  {2006})}\BibitemShut {NoStop}%
\bibitem [{\citenamefont {Nagaosa}\ \emph {et~al.}(2010)\citenamefont
  {Nagaosa}, \citenamefont {Sinova}, \citenamefont {Onoda}, \citenamefont
  {MacDonald},\ and\ \citenamefont {Ong}}]{nagaosa2010anomalous}%
  \BibitemOpen
  \bibfield  {author} {\bibinfo {author} {\bibfnamefont {N.}~\bibnamefont
  {Nagaosa}}, \bibinfo {author} {\bibfnamefont {J.}~\bibnamefont {Sinova}},
  \bibinfo {author} {\bibfnamefont {S.}~\bibnamefont {Onoda}}, \bibinfo
  {author} {\bibfnamefont {A.~H.}\ \bibnamefont {MacDonald}},\ and\ \bibinfo
  {author} {\bibfnamefont {N.~P.}\ \bibnamefont {Ong}},\ }\bibfield  {title}
  {\bibinfo {title} {Anomalous \ce{Hall} effect},\ }\href@noop {} {\bibfield
  {journal} {\bibinfo  {journal} {Rev. Mod. Phys.}\ }\textbf
  {\bibinfo {volume} {82}},\ \bibinfo {pages} {1539} (\bibinfo {year}
  {2010})}\BibitemShut {NoStop}%
\bibitem [{\citenamefont {Tian}\ \emph {et~al.}(2009)\citenamefont {Tian},
  \citenamefont {Ye},\ and\ \citenamefont {Jin}}]{tian2009proper}%
  \BibitemOpen
  \bibfield  {author} {\bibinfo {author} {\bibfnamefont {Y.}~\bibnamefont
  {Tian}}, \bibinfo {author} {\bibfnamefont {L.}~\bibnamefont {Ye}},\ and\
  \bibinfo {author} {\bibfnamefont {X.}~\bibnamefont {Jin}},\ }\bibfield
  {title} {\bibinfo {title} {Proper scaling of the anomalous \ce{Hall}
  effect},\ }\href@noop {} {\bibfield  {journal} {\bibinfo  {journal} {Phys. Rev. Lett.}\ }\textbf {\bibinfo {volume} {103}},\ \bibinfo {pages}
  {087206} (\bibinfo {year} {2009})}\BibitemShut {NoStop}%
\bibitem [{\citenamefont {Yue}\ and\ \citenamefont
  {Jin}(2017)}]{yue2017towards}%
  \BibitemOpen
  \bibfield  {author} {\bibinfo {author} {\bibfnamefont {D.}~\bibnamefont
  {Yue}}\ and\ \bibinfo {author} {\bibfnamefont {X.}~\bibnamefont {Jin}},\
  }\bibfield  {title} {\bibinfo {title} {Towards a better understanding of the
  anomalous \ce{Hall} effect},\ }\href@noop {} {\bibfield  {journal} {\bibinfo
  {journal} {J. Phys. Soc. Jpn.}\ }\textbf {\bibinfo
  {volume} {86}},\ \bibinfo {pages} {011006} (\bibinfo {year}
  {2017})}\BibitemShut {NoStop}%
\bibitem [{\citenamefont {Nakatsuji}\ \emph {et~al.}(2015)\citenamefont
  {Nakatsuji}, \citenamefont {Kiyohara},\ and\ \citenamefont
  {Higo}}]{nakatsuji2015large}%
  \BibitemOpen
  \bibfield  {author} {\bibinfo {author} {\bibfnamefont {S.}~\bibnamefont
  {Nakatsuji}}, \bibinfo {author} {\bibfnamefont {N.}~\bibnamefont
  {Kiyohara}},\ and\ \bibinfo {author} {\bibfnamefont {T.}~\bibnamefont
  {Higo}},\ }\bibfield  {title} {\bibinfo {title} {Large anomalous \ce{Hall}
  effect in a non-collinear antiferromagnet at room temperature},\ }\href@noop
  {} {\bibfield  {journal} {\bibinfo  {journal} {Nature}\ }\textbf {\bibinfo
  {volume} {527}},\ \bibinfo {pages} {212-215} (\bibinfo {year}
  {2015})}\BibitemShut {NoStop}%
\bibitem [{\citenamefont {Hao}\ \emph {et~al.}(2017)\citenamefont {Hao},
  \citenamefont {Chen}, \citenamefont {Wang},\ and\ \citenamefont
  {Xiao}}]{hao2017anomalous}%
  \BibitemOpen
  \bibfield  {author} {\bibinfo {author} {\bibfnamefont {Q.}~\bibnamefont
  {Hao}}, \bibinfo {author} {\bibfnamefont {W.}~\bibnamefont {Chen}}, \bibinfo
  {author} {\bibfnamefont {S.}~\bibnamefont {Wang}},\ and\ \bibinfo {author}
  {\bibfnamefont {G.}~\bibnamefont {Xiao}},\ }\bibfield  {title} {\bibinfo
  {title} {Anomalous \ce{Hall} effect and magnetic properties of
  \ce{Fe$_x$Pt$_{100-x}$} alloys with strong spin-orbit interaction},\
  }\href@noop {} {\bibfield  {journal} {\bibinfo  {journal} {J. Appl. Phys.}\ }\textbf {\bibinfo {volume} {122}},\ \bibinfo {pages} {033901}
  (\bibinfo {year} {2017})}\BibitemShut {NoStop}%
\bibitem [{\citenamefont {Smit}(1955)}]{smit1955spontaneous}%
  \BibitemOpen
  \bibfield  {author} {\bibinfo {author} {\bibfnamefont {J.}~\bibnamefont
  {Smit}},\ }\bibfield  {title} {\bibinfo {title} {The spontaneous \ce{Hall}
  effect in ferromagnetics \ce{I}},\ }\href@noop {} {\bibfield  {journal}
  {\bibinfo  {journal} {Physica}\ }\textbf {\bibinfo {volume} {21}},\ \bibinfo
  {pages} {877} (\bibinfo {year} {1955})}\BibitemShut {NoStop}%
\bibitem [{\citenamefont {Smit}(1958)}]{smit1958spontaneous}%
  \BibitemOpen
  \bibfield  {author} {\bibinfo {author} {\bibfnamefont {J.}~\bibnamefont
  {Smit}},\ }\bibfield  {title} {\bibinfo {title} {The spontaneous \ce{Hall}
  effect in ferromagnetics \ce{II}},\ }\href@noop {} {\bibfield  {journal}
  {\bibinfo  {journal} {Physica}\ }\textbf {\bibinfo {volume} {24}},\ \bibinfo
  {pages} {39} (\bibinfo {year} {1958})}\BibitemShut {NoStop}%
\bibitem [{\citenamefont {Karplus}\ and\ \citenamefont
  {Luttinger}(1954)}]{karplus1954hall}%
  \BibitemOpen
  \bibfield  {author} {\bibinfo {author} {\bibfnamefont {R.}~\bibnamefont
  {Karplus}}\ and\ \bibinfo {author} {\bibfnamefont {J.}~\bibnamefont
  {Luttinger}},\ }\bibfield  {title} {\bibinfo {title} {\ce{Hall} effect in
  ferromagnetics},\ }\href@noop {} {\bibfield  {journal} {\bibinfo  {journal}
  {Phys. Rev.}\ }\textbf {\bibinfo {volume} {95}},\ \bibinfo {pages}
  {1154} (\bibinfo {year} {1954})}\BibitemShut {NoStop}%
\bibitem [{\citenamefont {Sundaram}\ and\ \citenamefont
  {Niu}(1999)}]{sundaram1999wave}%
  \BibitemOpen
  \bibfield  {author} {\bibinfo {author} {\bibfnamefont {G.}~\bibnamefont
  {Sundaram}}\ and\ \bibinfo {author} {\bibfnamefont {Q.}~\bibnamefont {Niu}},\
  }\bibfield  {title} {\bibinfo {title} {Wave-packet dynamics in slowly
  perturbed crystals: Gradient corrections and \ce{Berry}-phase effects},\
  }\href@noop {} {\bibfield  {journal} {\bibinfo  {journal} {Phys. Rev. B}\ }\textbf {\bibinfo {volume} {59}},\ \bibinfo {pages} {14915} (\bibinfo
  {year} {1999})}\BibitemShut {NoStop}%
\bibitem [{\citenamefont {Xiao}\ \emph {et~al.}(2010)\citenamefont {Xiao},
  \citenamefont {Chang},\ and\ \citenamefont {Niu}}]{xiao2010berry}%
  \BibitemOpen
  \bibfield  {author} {\bibinfo {author} {\bibfnamefont {D.}~\bibnamefont
  {Xiao}}, \bibinfo {author} {\bibfnamefont {M.-C.}\ \bibnamefont {Chang}},\
  and\ \bibinfo {author} {\bibfnamefont {Q.}~\bibnamefont {Niu}},\ }\bibfield
  {title} {\bibinfo {title} {\ce{Berry} phase effects on electronic
  properties},\ }\href@noop {} {\bibfield  {journal} {\bibinfo  {journal}
  {Rev. Mod. Phys.}\ }\textbf {\bibinfo {volume} {82}},\ \bibinfo
  {pages} {1959} (\bibinfo {year} {2010})}\BibitemShut {NoStop}%
\bibitem [{\citenamefont {Manna}\ \emph
  {et~al.}(2018{\natexlab{a}})\citenamefont {Manna}, \citenamefont {Sun},
  \citenamefont {Muechler}, \citenamefont {K{\"u}bler},\ and\ \citenamefont
  {Felser}}]{manna2018heusler}%
  \BibitemOpen
  \bibfield  {author} {\bibinfo {author} {\bibfnamefont {K.}~\bibnamefont
  {Manna}}, \bibinfo {author} {\bibfnamefont {Y.}~\bibnamefont {Sun}}, \bibinfo
  {author} {\bibfnamefont {L.}~\bibnamefont {Muechler}}, \bibinfo {author}
  {\bibfnamefont {J.}~\bibnamefont {K{\"u}bler}},\ and\ \bibinfo {author}
  {\bibfnamefont {C.}~\bibnamefont {Felser}},\ }\bibfield  {title} {\bibinfo
  {title} {\ce{Heusler}, \ce{Weyl} and \ce{Berry}},\ }\href@noop {} {\bibfield
  {journal} {\bibinfo  {journal} { Nat Rev Mater}\ }\textbf {\bibinfo
  {volume} {3}},\ \bibinfo {pages} {244} (\bibinfo {year}
  {2018}{\natexlab{a}})}\BibitemShut {NoStop}%
\bibitem [{\citenamefont {Ye}\ \emph {et~al.}(1999)\citenamefont {Ye},
  \citenamefont {Kim}, \citenamefont {Millis}, \citenamefont {Shraiman},
  \citenamefont {Majumdar},\ and\ \citenamefont
  {Te{\v{s}}anovi{\'c}}}]{ye1999berry}%
  \BibitemOpen
  \bibfield  {author} {\bibinfo {author} {\bibfnamefont {J.}~\bibnamefont
  {Ye}}, \bibinfo {author} {\bibfnamefont {Y.~B.}\ \bibnamefont {Kim}},
  \bibinfo {author} {\bibfnamefont {A.}~\bibnamefont {Millis}}, \bibinfo
  {author} {\bibfnamefont {B.}~\bibnamefont {Shraiman}}, \bibinfo {author}
  {\bibfnamefont {P.}~\bibnamefont {Majumdar}},\ and\ \bibinfo {author}
  {\bibfnamefont {Z.}~\bibnamefont {Te{\v{s}}anovi{\'c}}},\ }\bibfield  {title}
  {\bibinfo {title} {\ce{Berry} phase theory of the anomalous \ce{Hall} effect:
  Application to colossal magnetoresistance manganites},\ }\href@noop {}
  {\bibfield  {journal} {\bibinfo  {journal} {Phys. Rev. Lett.}\
  }\textbf {\bibinfo {volume} {83}},\ \bibinfo {pages} {3737} (\bibinfo {year}
  {1999})}\BibitemShut {NoStop}%
\bibitem [{\citenamefont {He}\ \emph {et~al.}(2012)\citenamefont {He},
  \citenamefont {Moore},\ and\ \citenamefont {Varma}}]{he2012berry}%
  \BibitemOpen
  \bibfield  {author} {\bibinfo {author} {\bibfnamefont {Y.}~\bibnamefont
  {He}}, \bibinfo {author} {\bibfnamefont {J.}~\bibnamefont {Moore}},\ and\
  \bibinfo {author} {\bibfnamefont {C.}~\bibnamefont {Varma}},\ }\bibfield
  {title} {\bibinfo {title} {\ce{Berry} phase and anomalous \ce{Hall} effect in
  a three-orbital tight-binding hamiltonian},\ }\href@noop {} {\bibfield
  {journal} {\bibinfo  {journal} {Phys. Rev. B}\ }\textbf {\bibinfo
  {volume} {85}},\ \bibinfo {pages} {155106} (\bibinfo {year}
  {2012})}\BibitemShut {NoStop}%
\bibitem [{\citenamefont {Shen}\ \emph {et~al.}(2020)\citenamefont {Shen},
  \citenamefont {Yao}, \citenamefont {Zeng}, \citenamefont {Sun}, \citenamefont
  {Xi}, \citenamefont {Wu}, \citenamefont {Wang}, \citenamefont {Shen},
  \citenamefont {Liu},\ and\ \citenamefont {Liu}}]{shen2020local}%
  \BibitemOpen
  \bibfield  {author} {\bibinfo {author} {\bibfnamefont {J.}~\bibnamefont
  {Shen}}, \bibinfo {author} {\bibfnamefont {Q.}~\bibnamefont {Yao}}, \bibinfo
  {author} {\bibfnamefont {Q.}~\bibnamefont {Zeng}}, \bibinfo {author}
  {\bibfnamefont {H.}~\bibnamefont {Sun}}, \bibinfo {author} {\bibfnamefont
  {X.}~\bibnamefont {Xi}}, \bibinfo {author} {\bibfnamefont {G.}~\bibnamefont
  {Wu}}, \bibinfo {author} {\bibfnamefont {W.}~\bibnamefont {Wang}}, \bibinfo
  {author} {\bibfnamefont {B.}~\bibnamefont {Shen}}, \bibinfo {author}
  {\bibfnamefont {Q.}~\bibnamefont {Liu}},\ and\ \bibinfo {author}
  {\bibfnamefont {E.}~\bibnamefont {Liu}},\ }\bibfield  {title} {\bibinfo
  {title} {Local disorder-induced elevation of intrinsic anomalous \ce{Hall}
  conductance in an electron-doped magnetic \ce{Weyl} semimetal},\ }\href@noop
  {} {\bibfield  {journal} {\bibinfo  {journal} {Phys. Rev. Lett.}\
  }\textbf {\bibinfo {volume} {125}},\ \bibinfo {pages} {086602} (\bibinfo
  {year} {2020})}\BibitemShut {NoStop}%
\bibitem [{\citenamefont {Vidal}\ \emph {et~al.}(2011)\citenamefont {Vidal},
  \citenamefont {Schneider},\ and\ \citenamefont {Jakob}}]{vidal2011influence}%
  \BibitemOpen
  \bibfield  {author} {\bibinfo {author} {\bibfnamefont {E.~V.}\ \bibnamefont
  {Vidal}}, \bibinfo {author} {\bibfnamefont {H.}~\bibnamefont {Schneider}},\
  and\ \bibinfo {author} {\bibfnamefont {G.}~\bibnamefont {Jakob}},\ }\bibfield
   {title} {\bibinfo {title} {Influence of disorder on anomalous \ce{Hall}
  effect for \ce{Heusler} compounds},\ }\href@noop {} {\bibfield  {journal}
  {\bibinfo  {journal} {Phys. Rev. B}\ }\textbf {\bibinfo {volume} {83}},\
  \bibinfo {pages} {174410} (\bibinfo {year} {2011})}\BibitemShut {NoStop}%
\bibitem [{\citenamefont {Ernst}\ \emph {et~al.}(2019)\citenamefont {Ernst},
  \citenamefont {Sahoo}, \citenamefont {Sun}, \citenamefont {Nayak},
  \citenamefont {M\"uchler}, \citenamefont {Nayak}, \citenamefont {Kumar},
  \citenamefont {Gayles}, \citenamefont {Markou}, \citenamefont {Fecher},\ and\
  \citenamefont {Felser}}]{ernst2019anomalous}%
  \BibitemOpen
  \bibfield  {author} {\bibinfo {author} {\bibfnamefont {B.}~\bibnamefont
  {Ernst}}, \bibinfo {author} {\bibfnamefont {R.}~\bibnamefont {Sahoo}},
  \bibinfo {author} {\bibfnamefont {Y.}~\bibnamefont {Sun}}, \bibinfo {author}
  {\bibfnamefont {J.}~\bibnamefont {Nayak}}, \bibinfo {author} {\bibfnamefont
  {L.}~\bibnamefont {M\"uchler}}, \bibinfo {author} {\bibfnamefont {A.~K.}\
  \bibnamefont {Nayak}}, \bibinfo {author} {\bibfnamefont {N.}~\bibnamefont
  {Kumar}}, \bibinfo {author} {\bibfnamefont {J.}~\bibnamefont {Gayles}},
  \bibinfo {author} {\bibfnamefont {A.}~\bibnamefont {Markou}}, \bibinfo
  {author} {\bibfnamefont {G.~H.}\ \bibnamefont {Fecher}},\ and\ \bibinfo
  {author} {\bibfnamefont {C.}~\bibnamefont {Felser}},\ }\bibfield  {title}
  {\bibinfo {title} {Anomalous Hall effect and the role of Berry curvature in Co$_2$TiSn Heusler films},\ }\href
  {https://doi.org/10.1103/PhysRevB.100.054445} {\bibfield  {journal} {\bibinfo
   {journal} {Phys. Rev. B}\ }\textbf {\bibinfo {volume} {100}},\ \bibinfo
  {pages} {054445} (\bibinfo {year} {2019})}\BibitemShut {NoStop}%
\bibitem [{\citenamefont {Sakuraba}\ \emph {et~al.}(2020)\citenamefont
  {Sakuraba}, \citenamefont {Hyodo}, \citenamefont {Sakuma},\ and\
  \citenamefont {Mitani}}]{sakuraba2020giant}%
  \BibitemOpen
  \bibfield  {author} {\bibinfo {author} {\bibfnamefont {Y.}~\bibnamefont
  {Sakuraba}}, \bibinfo {author} {\bibfnamefont {K.}~\bibnamefont {Hyodo}},
  \bibinfo {author} {\bibfnamefont {A.}~\bibnamefont {Sakuma}},\ and\ \bibinfo
  {author} {\bibfnamefont {S.}~\bibnamefont {Mitani}},\ }\bibfield  {title}
  {\bibinfo {title} {Giant anomalous nernst effect in the
  \ce{Co$_2$MnAl$_{1-x}$Si$_ x$} \ce{Heusler} alloy induced by \ce{Fermi} level
  tuning and atomic ordering},\ }\href@noop {} {\bibfield  {journal} {\bibinfo
  {journal} {Phys. Rev. B}\ }\textbf {\bibinfo {volume} {101}},\ \bibinfo
  {pages} {134407} (\bibinfo {year} {2020})}\BibitemShut {NoStop}%
\bibitem [{\citenamefont {Kudrnovsk{\`y}}\ \emph {et~al.}(2013)\citenamefont
  {Kudrnovsk{\`y}}, \citenamefont {Drchal},\ and\ \citenamefont
  {Turek}}]{kudrnovsky2013anomalous}%
  \BibitemOpen
  \bibfield  {author} {\bibinfo {author} {\bibfnamefont {J.}~\bibnamefont
  {Kudrnovsk{\`y}}}, \bibinfo {author} {\bibfnamefont {V.}~\bibnamefont
  {Drchal}},\ and\ \bibinfo {author} {\bibfnamefont {I.}~\bibnamefont
  {Turek}},\ }\bibfield  {title} {\bibinfo {title} {Anomalous \ce{Hall} effect
  in stoichiometric \ce{Heusler} alloys with native disorder: A
  first-principles study},\ }\href@noop {} {\bibfield  {journal} {\bibinfo
  {journal} {Phys. Rev. B}\ }\textbf {\bibinfo {volume} {88}},\ \bibinfo
  {pages} {014422} (\bibinfo {year} {2013})}\BibitemShut {NoStop}%
\bibitem [{\citenamefont {Mende}\ \emph {et~al.}(2021)\citenamefont {Mende},
  \citenamefont {Noky}, \citenamefont {Guin}, \citenamefont {Fecher},
  \citenamefont {Manna}, \citenamefont {Adler}, \citenamefont {Schnelle},
  \citenamefont {Sun}, \citenamefont {Fu},\ and\ \citenamefont
  {Felser}}]{mende2021large}%
  \BibitemOpen
  \bibfield  {author} {\bibinfo {author} {\bibfnamefont {F.}~\bibnamefont
  {Mende}}, \bibinfo {author} {\bibfnamefont {J.}~\bibnamefont {Noky}},
  \bibinfo {author} {\bibfnamefont {S.~N.}\ \bibnamefont {Guin}}, \bibinfo
  {author} {\bibfnamefont {G.~H.}\ \bibnamefont {Fecher}}, \bibinfo {author}
  {\bibfnamefont {K.}~\bibnamefont {Manna}}, \bibinfo {author} {\bibfnamefont
  {P.}~\bibnamefont {Adler}}, \bibinfo {author} {\bibfnamefont
  {W.}~\bibnamefont {Schnelle}}, \bibinfo {author} {\bibfnamefont
  {Y.}~\bibnamefont {Sun}}, \bibinfo {author} {\bibfnamefont {C.}~\bibnamefont
  {Fu}},\ and\ \bibinfo {author} {\bibfnamefont {C.}~\bibnamefont {Felser}},\
  }\bibfield  {title} {\bibinfo {title} {Large anomalous \ce{Hall} and
  \ce{Nernst} effects in high curie-temperature iron-based \ce{Heusler}
  compounds},\ }\href@noop {} {\bibfield  {journal} {\bibinfo  {journal}
  { Adv. Sci.}\ \ }\textbf {\bibinfo {volume} {8}},\ \bibinfo
  {pages} {2100782} (\bibinfo {year}
  {2021})}\BibitemShut {NoStop}%
\bibitem [{\citenamefont {Brown}\ \emph {et~al.}(2000)\citenamefont {Brown},
  \citenamefont {Neumann}, \citenamefont {Webster},\ and\ \citenamefont
  {Ziebeck}}]{brown2000magnetization}%
  \BibitemOpen
  \bibfield  {author} {\bibinfo {author} {\bibfnamefont {P.}~\bibnamefont
  {Brown}}, \bibinfo {author} {\bibfnamefont {K.-U.}\ \bibnamefont {Neumann}},
  \bibinfo {author} {\bibfnamefont {P.}~\bibnamefont {Webster}},\ and\ \bibinfo
  {author} {\bibfnamefont {K.}~\bibnamefont {Ziebeck}},\ }\bibfield  {title}
  {\bibinfo {title} {The magnetization distributions in some \ce{Heusler}
  alloys proposed as half-metallic ferromagnets},\ }\href@noop {} {\bibfield
  {journal} {\bibinfo  {journal} {J. Phys.: Condens. Matter}\
  }\textbf {\bibinfo {volume} {12}},\ \bibinfo {pages} {1827} (\bibinfo {year}
  {2000})}\BibitemShut {NoStop}%
\bibitem [{\citenamefont {Galanakis}\ \emph {et~al.}(2002)\citenamefont
  {Galanakis}, \citenamefont {Dederichs},\ and\ \citenamefont
  {Papanikolaou}}]{galanakis2002slater}%
  \BibitemOpen
  \bibfield  {author} {\bibinfo {author} {\bibfnamefont {I.}~\bibnamefont
  {Galanakis}}, \bibinfo {author} {\bibfnamefont {P.}~\bibnamefont
  {Dederichs}},\ and\ \bibinfo {author} {\bibfnamefont {N.}~\bibnamefont
  {Papanikolaou}},\ }\bibfield  {title} {\bibinfo {title} {Slater-pauling
  behavior and origin of the half-metallicity of the full-\ce{Heusler}
  alloys},\ }\href@noop {} {\bibfield  {journal} {\bibinfo  {journal} {Phys. Rev. B}\ }\textbf {\bibinfo {volume} {66}},\ \bibinfo {pages} {174429}
  (\bibinfo {year} {2002})}\BibitemShut {NoStop}%
\bibitem [{\citenamefont {K{\"u}bler}\ \emph {et~al.}(2007)\citenamefont
  {K{\"u}bler}, \citenamefont {Fecher},\ and\ \citenamefont
  {Felser}}]{kubler2007understanding}%
  \BibitemOpen
  \bibfield  {author} {\bibinfo {author} {\bibfnamefont {J.}~\bibnamefont
  {K{\"u}bler}}, \bibinfo {author} {\bibfnamefont {G.}~\bibnamefont {Fecher}},\
  and\ \bibinfo {author} {\bibfnamefont {C.}~\bibnamefont {Felser}},\
  }\bibfield  {title} {\bibinfo {title} {Understanding the trend in the
  \ce{Curie} temperatures of \ce{Co$_2$}-based \ce{Heusler} compounds: Ab initio
  calculations},\ }\href@noop {} {\bibfield  {journal} {\bibinfo  {journal}
  {Phys. Rev. B}\ }\textbf {\bibinfo {volume} {76}},\ \bibinfo {pages}
  {024414} (\bibinfo {year} {2007})}\BibitemShut {NoStop}%
\bibitem [{\citenamefont {Kandpal}\ \emph {et~al.}(2007)\citenamefont
  {Kandpal}, \citenamefont {Fecher},\ and\ \citenamefont
  {Felser}}]{kandpal2007calculated}%
  \BibitemOpen
  \bibfield  {author} {\bibinfo {author} {\bibfnamefont {H.~C.}\ \bibnamefont
  {Kandpal}}, \bibinfo {author} {\bibfnamefont {G.~H.}\ \bibnamefont
  {Fecher}},\ and\ \bibinfo {author} {\bibfnamefont {C.}~\bibnamefont
  {Felser}},\ }\bibfield  {title} {\bibinfo {title} {Calculated electronic and
  magnetic properties of the half-metallic, transition metal based \ce{Heusler}
  compounds},\ }\href@noop {} {\bibfield  {journal} {\bibinfo  {journal}
  {J. Phys. D: Appl. Phys.}\ }\textbf {\bibinfo {volume} {40}},\
  \bibinfo {pages} {1507} (\bibinfo {year} {2007})}\BibitemShut {NoStop}%
\bibitem [{\citenamefont {Roy}\ \emph {et~al.}(2020)\citenamefont {Roy},
  \citenamefont {Singha}, \citenamefont {Ghosh}, \citenamefont {Pariari},\ and\
  \citenamefont {Mandal}}]{roy2020anomalous}%
  \BibitemOpen
  \bibfield  {author} {\bibinfo {author} {\bibfnamefont {S.}~\bibnamefont
  {Roy}}, \bibinfo {author} {\bibfnamefont {R.}~\bibnamefont {Singha}},
  \bibinfo {author} {\bibfnamefont {A.}~\bibnamefont {Ghosh}}, \bibinfo
  {author} {\bibfnamefont {A.}~\bibnamefont {Pariari}},\ and\ \bibinfo {author}
  {\bibfnamefont {P.}~\bibnamefont {Mandal}},\ }\bibfield  {title} {\bibinfo
  {title} {Anomalous \ce{Hall} effect in the half-metallic \ce{Heusler}
  compound \ce{Co$_2$TiX (X= Si, Ge)}},\ }\href@noop {} {\bibfield  {journal}
  {\bibinfo  {journal} {Phys. Rev. B}\ }\textbf {\bibinfo {volume}
  {102}},\ \bibinfo {pages} {085147} (\bibinfo {year} {2020})}\BibitemShut
  {NoStop}%
\bibitem [{\citenamefont {Reichlova}\ \emph {et~al.}(2018)\citenamefont
  {Reichlova}, \citenamefont {Schlitz}, \citenamefont {Beckert}, \citenamefont
  {Swekis}, \citenamefont {Markou}, \citenamefont {Chen}, \citenamefont
  {Kriegner}, \citenamefont {Fabretti}, \citenamefont {Hyeon~Park},
  \citenamefont {Niemann} \emph {et~al.}}]{reichlova2018large}%
  \BibitemOpen
  \bibfield  {author} {\bibinfo {author} {\bibfnamefont {H.}~\bibnamefont
  {Reichlova}}, \bibinfo {author} {\bibfnamefont {R.}~\bibnamefont {Schlitz}},
  \bibinfo {author} {\bibfnamefont {S.}~\bibnamefont {Beckert}}, \bibinfo
  {author} {\bibfnamefont {P.}~\bibnamefont {Swekis}}, \bibinfo {author}
  {\bibfnamefont {A.}~\bibnamefont {Markou}}, \bibinfo {author} {\bibfnamefont
  {Y.-C.}\ \bibnamefont {Chen}}, \bibinfo {author} {\bibfnamefont
  {D.}~\bibnamefont {Kriegner}}, \bibinfo {author} {\bibfnamefont
  {S.}~\bibnamefont {Fabretti}}, \bibinfo {author} {\bibfnamefont
  {G.}~\bibnamefont {Hyeon~Park}}, \bibinfo {author} {\bibfnamefont
  {A.}~\bibnamefont {Niemann}}, \bibinfo {author} {\bibfnamefont
  {S.}~\bibnamefont {Sudheendra}}, \bibinfo {author} {\bibfnamefont
  {A.}~\bibnamefont {Thomas}}, \bibinfo {author} {\bibfnamefont
  {K.}~\bibnamefont {Nielsch}}, \bibinfo {author} {\bibfnamefont
  {C.}~\bibnamefont {Felser}},\ and\ \bibinfo {author} {\bibfnamefont
  {S. T. B.}~\bibnamefont {Goennenwein}},\ }\bibfield  {title} {\bibinfo
  {title} {Large anomalous Nernst effect in thin films of the Weyl semimetal
  Co$_2$MnGa},\ }\href@noop {} {\bibfield  {journal} {\bibinfo  {journal} {Appl. Phys. Lett.}\ }\textbf {\bibinfo {volume} {113}},\ \bibinfo {pages}
  {212405} (\bibinfo {year} {2018})}\BibitemShut {NoStop}%
\bibitem [{\citenamefont {Li}\ \emph {et~al.}(2020)\citenamefont {Li},
  \citenamefont {Koo}, \citenamefont {Ning}, \citenamefont {Li}, \citenamefont
  {Miao}, \citenamefont {Min}, \citenamefont {Zhu}, \citenamefont {Wang},
  \citenamefont {Alem}, \citenamefont {Liu} \emph {et~al.}}]{li2020giant}%
  \BibitemOpen
  \bibfield  {author} {\bibinfo {author} {\bibfnamefont {P.}~\bibnamefont
  {Li}}, \bibinfo {author} {\bibfnamefont {J.}~\bibnamefont {Koo}}, \bibinfo
  {author} {\bibfnamefont {W.}~\bibnamefont {Ning}}, \bibinfo {author}
  {\bibfnamefont {J.}~\bibnamefont {Li}}, \bibinfo {author} {\bibfnamefont
  {L.}~\bibnamefont {Miao}}, \bibinfo {author} {\bibfnamefont {L.}~\bibnamefont
  {Min}}, \bibinfo {author} {\bibfnamefont {Y.}~\bibnamefont {Zhu}}, \bibinfo
  {author} {\bibfnamefont {Y.}~\bibnamefont {Wang}}, \bibinfo {author}
  {\bibfnamefont {N.}~\bibnamefont {Alem}}, \bibinfo {author} {\bibfnamefont
  {C.-X.}\ \bibnamefont {Liu}}, \bibinfo {author} {\bibfnamefont
  {Z.}\ \bibnamefont {Mao}},\ and\ \bibinfo {author} {\bibfnamefont
  {B.}\ \bibnamefont {Yan}},\ }\bibfield  {title} {\bibinfo
  {title} {Giant room temperature anomalous \ce{Hall} effect and tunable topology in
  a ferromagnetic topological semimetal \ce{Co$_2$MnAl}},\ }\href@noop {}
  {\bibfield  {journal} {\bibinfo  {journal} {Nat. Commun.}\ }\textbf {\bibinfo
  {volume} {11}},\ \bibinfo {pages} {3476} (\bibinfo {year}
  {2020})}\BibitemShut {NoStop}%
\bibitem [{\citenamefont {Guin}\ \emph {et~al.}(2019)\citenamefont {Guin},
  \citenamefont {Manna}, \citenamefont {Noky}, \citenamefont {Watzman},
  \citenamefont {Fu}, \citenamefont {Kumar}, \citenamefont {Schnelle},
  \citenamefont {Shekhar}, \citenamefont {Sun}, \citenamefont {Gooth} \emph
  {et~al.}}]{guin2019anomalous}%
  \BibitemOpen
  \bibfield  {author} {\bibinfo {author} {\bibfnamefont {S.~N.}\ \bibnamefont
  {Guin}}, \bibinfo {author} {\bibfnamefont {K.}~\bibnamefont {Manna}},
  \bibinfo {author} {\bibfnamefont {J.}~\bibnamefont {Noky}}, \bibinfo {author}
  {\bibfnamefont {S.~J.}\ \bibnamefont {Watzman}}, \bibinfo {author}
  {\bibfnamefont {C.}~\bibnamefont {Fu}}, \bibinfo {author} {\bibfnamefont
  {N.}~\bibnamefont {Kumar}}, \bibinfo {author} {\bibfnamefont
  {W.}~\bibnamefont {Schnelle}}, \bibinfo {author} {\bibfnamefont
  {C.}~\bibnamefont {Shekhar}}, \bibinfo {author} {\bibfnamefont
  {Y.}~\bibnamefont {Sun}}, \bibinfo {author} {\bibfnamefont {J.}~\bibnamefont
  {Gooth}},\ and\  \bibinfo {author} {\bibfnamefont {C.}~\bibnamefont
  {Felser}},\ }\bibfield  {title} {\bibinfo {title} {Anomalous
  Nernst effect beyond the magnetization scaling relation in the ferromagnetic
  Heusler compound Co$_2$MnGa},\ }\href@noop {} {\bibfield  {journal} {\bibinfo
  {journal} {NPG Asia Mater.}\ }\textbf {\bibinfo {volume} {11}},\ \bibinfo
  {pages} {16} (\bibinfo {year} {2019})}\BibitemShut {NoStop}%
\bibitem [{\citenamefont {Husain}\ \emph {et~al.}(2019)\citenamefont {Husain},
  \citenamefont {Sisodia}, \citenamefont {Chaurasiya}, \citenamefont {Kumar},
  \citenamefont {Singh}, \citenamefont {Yadav}, \citenamefont {Akansel},
  \citenamefont {Chae}, \citenamefont {Barman}, \citenamefont {Muduli} \emph
  {et~al.}}]{husain2019observation}%
  \BibitemOpen
  \bibfield  {author} {\bibinfo {author} {\bibfnamefont {S.}~\bibnamefont
  {Husain}}, \bibinfo {author} {\bibfnamefont {N.}~\bibnamefont {Sisodia}},
  \bibinfo {author} {\bibfnamefont {A.~K.}\ \bibnamefont {Chaurasiya}},
  \bibinfo {author} {\bibfnamefont {A.}~\bibnamefont {Kumar}}, \bibinfo
  {author} {\bibfnamefont {J.~P.}\ \bibnamefont {Singh}}, \bibinfo {author}
  {\bibfnamefont {B.~S.}\ \bibnamefont {Yadav}}, \bibinfo {author}
  {\bibfnamefont {S.}~\bibnamefont {Akansel}}, \bibinfo {author} {\bibfnamefont
  {K.~H.}\ \bibnamefont {Chae}}, \bibinfo {author} {\bibfnamefont
  {A.}~\bibnamefont {Barman}}, \bibinfo {author} {\bibfnamefont
  {P.}~\bibnamefont {Muduli}}, \bibinfo {author} {\bibfnamefont
  {P.}~\bibnamefont {Svedlindh}},\ and\ \bibinfo {author} {\bibfnamefont
  {S.}~\bibnamefont {Chaudhary}},\ }\bibfield  {title} {\bibinfo
  {title} {Observation of skyrmions at room temperature in \ce{Co$_2$FeAl}
  \ce{Heusler} alloy ultrathin film heterostructures},\ }\href@noop {}
  {\bibfield  {journal} {\bibinfo  {journal} {Sci. Rep.}\ }\textbf {\bibinfo
  {volume} {9}},\ \bibinfo {pages} {1085} (\bibinfo {year} {2019})}\BibitemShut
  {NoStop}%
\bibitem [{\citenamefont {Zhang}\ \emph {et~al.}(2018)\citenamefont {Zhang},
  \citenamefont {Xu}, \citenamefont {Lai}, \citenamefont {Lu}, \citenamefont
  {Lu}, \citenamefont {Chen}, \citenamefont {Niu}, \citenamefont {Gu},
  \citenamefont {Liu}, \citenamefont {Wang} \emph {et~al.}}]{zhang2018direct}%
  \BibitemOpen
  \bibfield  {author} {\bibinfo {author} {\bibfnamefont {X.}~\bibnamefont
  {Zhang}}, \bibinfo {author} {\bibfnamefont {H.}~\bibnamefont {Xu}}, \bibinfo
  {author} {\bibfnamefont {B.}~\bibnamefont {Lai}}, \bibinfo {author}
  {\bibfnamefont {Q.}~\bibnamefont {Lu}}, \bibinfo {author} {\bibfnamefont
  {X.}~\bibnamefont {Lu}}, \bibinfo {author} {\bibfnamefont {Y.}~\bibnamefont
  {Chen}}, \bibinfo {author} {\bibfnamefont {W.}~\bibnamefont {Niu}}, \bibinfo
  {author} {\bibfnamefont {C.}~\bibnamefont {Gu}}, \bibinfo {author}
  {\bibfnamefont {W.}~\bibnamefont {Liu}}, \bibinfo {author} {\bibfnamefont
  {X.}~\bibnamefont {Wang}}, \bibinfo {author} {\bibfnamefont
  {C.}~\bibnamefont {Liu}}, \bibinfo {author} {\bibfnamefont
  {Y.}~\bibnamefont {Nie}}, \bibinfo {author} {\bibfnamefont
  {L.}~\bibnamefont {He}},\ and\ \bibinfo {author} {\bibfnamefont
  {Y.}~\bibnamefont {Hu}},\ }\bibfield  {title} {\bibinfo
  {title} {Direct observation of high spin polarization in \ce{Co$_2$FeAl} thin
  films},\ }\href@noop {} {\bibfield  {journal} {\bibinfo  {journal} {Sci. Rep.}\
  }\textbf {\bibinfo {volume} {8}},\ \bibinfo {pages} {8074} (\bibinfo {year}
  {2018})}\BibitemShut {NoStop}%
\bibitem [{\citenamefont {Husain}\ \emph {et~al.}(2016)\citenamefont {Husain},
  \citenamefont {Akansel}, \citenamefont {Kumar}, \citenamefont {Svedlindh},\
  and\ \citenamefont {Chaudhary}}]{husain2016growth}%
  \BibitemOpen
  \bibfield  {author} {\bibinfo {author} {\bibfnamefont {S.}~\bibnamefont
  {Husain}}, \bibinfo {author} {\bibfnamefont {S.}~\bibnamefont {Akansel}},
  \bibinfo {author} {\bibfnamefont {A.}~\bibnamefont {Kumar}}, \bibinfo
  {author} {\bibfnamefont {P.}~\bibnamefont {Svedlindh}},\ and\ \bibinfo
  {author} {\bibfnamefont {S.}~\bibnamefont {Chaudhary}},\ }\bibfield  {title}
  {\bibinfo {title} {Growth of \ce{Co$_2$FeAl} \ce{Heusler} alloy thin films on
  \ce{Si} (100) having very small gilbert damping by ion beam sputtering},\
  }\href@noop {} {\bibfield  {journal} {\bibinfo  {journal} {Sci. Rep.}\ }\textbf {\bibinfo {volume} {6}},\ \bibinfo {pages} {28692}
  (\bibinfo {year} {2016})}\BibitemShut {NoStop}%
\bibitem [{\citenamefont {Malik}\ \emph {et~al.}(2021)\citenamefont {Malik},
  \citenamefont {Delczeg-Czirjak}, \citenamefont {Knut}, \citenamefont
  {Thonig}, \citenamefont {Vaskivskyi}, \citenamefont {Phuyal}, \citenamefont
  {Gupta}, \citenamefont {Jana}, \citenamefont {Stefanuik}, \citenamefont
  {Kvashnin} \emph {et~al.}}]{malik2021ultrafast}%
  \BibitemOpen
  \bibfield  {author} {\bibinfo {author} {\bibfnamefont {R.}~\bibnamefont
  {Malik}}, \bibinfo {author} {\bibfnamefont {E.}~\bibnamefont
  {Delczeg-Czirjak}}, \bibinfo {author} {\bibfnamefont {R.}~\bibnamefont
  {Knut}}, \bibinfo {author} {\bibfnamefont {D.}~\bibnamefont {Thonig}},
  \bibinfo {author} {\bibfnamefont {I.}~\bibnamefont {Vaskivskyi}}, \bibinfo
  {author} {\bibfnamefont {D.}~\bibnamefont {Phuyal}}, \bibinfo {author}
  {\bibfnamefont {R.}~\bibnamefont {Gupta}}, \bibinfo {author} {\bibfnamefont
  {S.}~\bibnamefont {Jana}}, \bibinfo {author} {\bibfnamefont {R.}~\bibnamefont
  {Stefanuik}}, \bibinfo {author} {\bibfnamefont {Y.}~\bibnamefont {Kvashnin}},\bibinfo {author} {\bibfnamefont {S.}~\bibnamefont {Husain}},\bibinfo {author} {\bibfnamefont {A.}~\bibnamefont {Kumar}},\bibinfo {author} {\bibfnamefont {Y.}~\bibnamefont {Svedlindh}},\bibinfo {author} {\bibfnamefont {J.}~\bibnamefont {Söderström}},\bibinfo {author}{\bibfnamefont {O.}~\bibnamefont { Eriksson}},\ and\ \bibinfo {author} {\bibfnamefont {O.}~\bibnamefont {Karis}},\ }\bibfield  {title} {\bibinfo {title} {Ultrafast
  magnetization dynamics in the half-metallic \ce{Heusler} alloy
  \ce{Co$_2$FeAl}},\ }\href@noop {} {\bibfield  {journal} {\bibinfo  {journal}
  {Phys. Rev. B}\ }\textbf {\bibinfo {volume} {104}},\ \bibinfo {pages}
  {L100408} (\bibinfo {year} {2021})}\BibitemShut {NoStop}%
\bibitem [{\citenamefont {Kumar}\ \emph {et~al.}(2017)\citenamefont {Kumar},
  \citenamefont {Pan}, \citenamefont {Husain}, \citenamefont {Akansel},
  \citenamefont {Brucas}, \citenamefont {Bergqvist}, \citenamefont
  {Chaudhary},\ and\ \citenamefont {Svedlindh}}]{PhysRevB.96.224425}%
  \BibitemOpen
  \bibfield  {author} {\bibinfo {author} {\bibfnamefont {A.}~\bibnamefont
  {Kumar}}, \bibinfo {author} {\bibfnamefont {F.}~\bibnamefont {Pan}}, \bibinfo
  {author} {\bibfnamefont {S.}~\bibnamefont {Husain}}, \bibinfo {author}
  {\bibfnamefont {S.}~\bibnamefont {Akansel}}, \bibinfo {author} {\bibfnamefont
  {R.}~\bibnamefont {Brucas}}, \bibinfo {author} {\bibfnamefont
  {L.}~\bibnamefont {Bergqvist}}, \bibinfo {author} {\bibfnamefont
  {S.}~\bibnamefont {Chaudhary}},\ and\ \bibinfo {author} {\bibfnamefont
  {P.}~\bibnamefont {Svedlindh}},\ }\bibfield  {title} {\bibinfo {title}
  {Temperature-dependent gilbert damping of Co$_2$FeAl
  thin films with different degree of atomic order},\ }\href
  {https://doi.org/10.1103/PhysRevB.96.224425} {\bibfield  {journal} {\bibinfo
  {journal} {Phys. Rev. B}\ }\textbf {\bibinfo {volume} {96}},\ \bibinfo
  {pages} {224425} (\bibinfo {year} {2017})}\BibitemShut {NoStop}%
\bibitem [{\citenamefont {Husain}\ \emph {et~al.}(2017)\citenamefont {Husain},
  \citenamefont {Kumar}, \citenamefont {Akansel}, \citenamefont {Svedlindh},\
  and\ \citenamefont {Chaudhary}}]{husain2017anomalous}%
  \BibitemOpen
  \bibfield  {author} {\bibinfo {author} {\bibfnamefont {S.}~\bibnamefont
  {Husain}}, \bibinfo {author} {\bibfnamefont {A.}~\bibnamefont {Kumar}},
  \bibinfo {author} {\bibfnamefont {S.}~\bibnamefont {Akansel}}, \bibinfo
  {author} {\bibfnamefont {P.}~\bibnamefont {Svedlindh}},\ and\ \bibinfo
  {author} {\bibfnamefont {S.}~\bibnamefont {Chaudhary}},\ }\bibfield  {title}
  {\bibinfo {title} {Anomalous \ce{Hall} effect in ion-beam sputtered
  \ce{Co$_2$FeAl} full \ce{Heusler} alloy thin films},\ }\href@noop {} {\bibfield
  {journal} {\bibinfo  {journal} {J. Magn. Magn. Mater.}\
  }\textbf {\bibinfo {volume} {442}},\ \bibinfo {pages} {288} (\bibinfo {year}
  {2017})}\BibitemShut {NoStop}%
\bibitem [{\citenamefont {Wang}\ \emph {et~al.}(2014)\citenamefont {Wang},
  \citenamefont {Li}, \citenamefont {Du}, \citenamefont {Dai}, \citenamefont
  {Liu}, \citenamefont {Liu}, \citenamefont {Liu}, \citenamefont {Wang},\ and\
  \citenamefont {Wu}}]{wang2014structural}%
  \BibitemOpen
  \bibfield  {author} {\bibinfo {author} {\bibfnamefont {X.}~\bibnamefont
  {Wang}}, \bibinfo {author} {\bibfnamefont {Y.}~\bibnamefont {Li}}, \bibinfo
  {author} {\bibfnamefont {Y.}~\bibnamefont {Du}}, \bibinfo {author}
  {\bibfnamefont {X.}~\bibnamefont {Dai}}, \bibinfo {author} {\bibfnamefont
  {G.}~\bibnamefont {Liu}}, \bibinfo {author} {\bibfnamefont {E.}~\bibnamefont
  {Liu}}, \bibinfo {author} {\bibfnamefont {Z.}~\bibnamefont {Liu}}, \bibinfo
  {author} {\bibfnamefont {W.}~\bibnamefont {Wang}},\ and\ \bibinfo {author}
  {\bibfnamefont {G.}~\bibnamefont {Wu}},\ }\bibfield  {title} {\bibinfo
  {title} {Structural, magnetic and transport properties of \ce{Co$_2$FeAl}
  \ce{Heusler} films with varying thickness},\ }\href@noop {} {\bibfield
  {journal} {\bibinfo  {journal} {J. Magn. Magn. Mater.}\
  }\textbf {\bibinfo {volume} {362}},\ \bibinfo {pages} {52} (\bibinfo {year}
  {2014})}\BibitemShut {NoStop}%
\bibitem [{\citenamefont {Kogachi}\ \emph {et~al.}(2006)\citenamefont
  {Kogachi}, \citenamefont {Tadachi},\ and\ \citenamefont
  {Nakanishi}}]{kogachi2006structural}%
  \BibitemOpen
  \bibfield  {author} {\bibinfo {author} {\bibfnamefont {M.}~\bibnamefont
  {Kogachi}}, \bibinfo {author} {\bibfnamefont {N.}~\bibnamefont {Tadachi}},\
  and\ \bibinfo {author} {\bibfnamefont {T.}~\bibnamefont {Nakanishi}},\
  }\bibfield  {title} {\bibinfo {title} {Structural properties and magnetic
  behavior in \ce{CoFe$_{1-x}$Al$_x$} alloys},\ }\href@noop {} {\bibfield
  {journal} {\bibinfo  {journal} {Intermetallics}\ }\textbf {\bibinfo {volume}
  {14}},\ \bibinfo {pages} {742} (\bibinfo {year} {2006})}\BibitemShut
  {NoStop}%
\bibitem [{\citenamefont {Imort}\ \emph {et~al.}(2012)\citenamefont {Imort},
  \citenamefont {Thomas}, \citenamefont {Reiss},\ and\ \citenamefont
  {Thomas}}]{imort2012anomalous}%
  \BibitemOpen
  \bibfield  {author} {\bibinfo {author} {\bibfnamefont {I.-M.}\ \bibnamefont
  {Imort}}, \bibinfo {author} {\bibfnamefont {P.}~\bibnamefont {Thomas}},
  \bibinfo {author} {\bibfnamefont {G.}~\bibnamefont {Reiss}},\ and\ \bibinfo
  {author} {\bibfnamefont {A.}~\bibnamefont {Thomas}},\ }\bibfield  {title}
  {\bibinfo {title} {Anomalous \ce{Hall} effect in the \ce{Co}-based \ce{Heusler}
  compounds \ce{Co$_2$FeSi} and \ce{Co$_2$FeAl}},\ }\href@noop {} {\bibfield
  {journal} {\bibinfo  {journal} {J.  Appl.  Phys}\ }\textbf
  {\bibinfo {volume} {111}},\ \bibinfo {pages} {07D313} (\bibinfo {year}
  {2012})}\BibitemShut {NoStop}%
\bibitem [{\citenamefont {Giannozzi}\ \emph {et~al.}(2009)\citenamefont
  {Giannozzi}, \citenamefont {Baroni}, \citenamefont {Bonini}, \citenamefont
  {Calandra}, \citenamefont {Car}, \citenamefont {Cavazzoni}, \citenamefont
  {Ceresoli}, \citenamefont {Chiarotti}, \citenamefont {Cococcioni},
  \citenamefont {Dabo}, \citenamefont {Corso}, \citenamefont {de~Gironcoli},
  \citenamefont {Fabris}, \citenamefont {Fratesi}, \citenamefont {Gebauer},
  \citenamefont {Gerstmann}, \citenamefont {Gougoussis}, \citenamefont
  {Kokalj}, \citenamefont {Lazzeri}, \citenamefont {Martin-Samos},
  \citenamefont {Marzari}, \citenamefont {Mauri}, \citenamefont {Mazzarello},
  \citenamefont {Paolini}, \citenamefont {Pasquarello}, \citenamefont
  {Paulatto}, \citenamefont {Sbraccia}, \citenamefont {Scandolo}, \citenamefont
  {Sclauzero}, \citenamefont {Seitsonen}, \citenamefont {Smogunov},
  \citenamefont {Umari},\ and\ \citenamefont
  {Wentzcovitch}}]{giannozzi2009quantum}%
  \BibitemOpen
  \bibfield  {author} {\bibinfo {author} {\bibfnamefont {P.}~\bibnamefont
  {Giannozzi}}, \bibinfo {author} {\bibfnamefont {S.}~\bibnamefont {Baroni}},
  \bibinfo {author} {\bibfnamefont {N.}~\bibnamefont {Bonini}}, \bibinfo
  {author} {\bibfnamefont {M.}~\bibnamefont {Calandra}}, \bibinfo {author}
  {\bibfnamefont {R.}~\bibnamefont {Car}}, \bibinfo {author} {\bibfnamefont
  {C.}~\bibnamefont {Cavazzoni}}, \bibinfo {author} {\bibfnamefont
  {D.}~\bibnamefont {Ceresoli}}, \bibinfo {author} {\bibfnamefont {G.~L.}\
  \bibnamefont {Chiarotti}}, \bibinfo {author} {\bibfnamefont {M.}~\bibnamefont
  {Cococcioni}}, \bibinfo {author} {\bibfnamefont {I.}~\bibnamefont {Dabo}},
  \bibinfo {author} {\bibfnamefont {A.~D.}\ \bibnamefont {Corso}}, \bibinfo
  {author} {\bibfnamefont {S.}~\bibnamefont {de~Gironcoli}}, \bibinfo {author}
  {\bibfnamefont {S.}~\bibnamefont {Fabris}}, \bibinfo {author} {\bibfnamefont
  {G.}~\bibnamefont {Fratesi}}, \bibinfo {author} {\bibfnamefont
  {R.}~\bibnamefont {Gebauer}}, \bibinfo {author} {\bibfnamefont
  {U.}~\bibnamefont {Gerstmann}}, \bibinfo {author} {\bibfnamefont
  {C.}~\bibnamefont {Gougoussis}}, \bibinfo {author} {\bibfnamefont
  {A.}~\bibnamefont {Kokalj}}, \bibinfo {author} {\bibfnamefont
  {M.}~\bibnamefont {Lazzeri}}, \bibinfo {author} {\bibfnamefont
  {L.}~\bibnamefont {Martin-Samos}}, \bibinfo {author} {\bibfnamefont
  {N.}~\bibnamefont {Marzari}}, \bibinfo {author} {\bibfnamefont
  {F.}~\bibnamefont {Mauri}}, \bibinfo {author} {\bibfnamefont
  {R.}~\bibnamefont {Mazzarello}}, \bibinfo {author} {\bibfnamefont
  {S.}~\bibnamefont {Paolini}}, \bibinfo {author} {\bibfnamefont
  {A.}~\bibnamefont {Pasquarello}}, \bibinfo {author} {\bibfnamefont
  {L.}~\bibnamefont {Paulatto}}, \bibinfo {author} {\bibfnamefont
  {C.}~\bibnamefont {Sbraccia}}, \bibinfo {author} {\bibfnamefont
  {S.}~\bibnamefont {Scandolo}}, \bibinfo {author} {\bibfnamefont
  {G.}~\bibnamefont {Sclauzero}}, \bibinfo {author} {\bibfnamefont {A.~P.}\
  \bibnamefont {Seitsonen}}, \bibinfo {author} {\bibfnamefont {A.}~\bibnamefont
  {Smogunov}}, \bibinfo {author} {\bibfnamefont {P.}~\bibnamefont {Umari}},\
  and\ \bibinfo {author} {\bibfnamefont {R.~M.}\ \bibnamefont {Wentzcovitch}},\
  }\bibfield  {title} {\bibinfo {title} {{QUANTUM} {ESPRESSO}: a modular and
  open-source software project for quantum simulations of materials},\ }\href
  {https://doi.org/10.1088/0953-8984/21/39/395502} {\bibfield  {journal}
  {\bibinfo  {journal} {J. Phys. Condens. Matter}\ }\textbf {\bibinfo {volume}
  {21}},\ \bibinfo {pages} {395502} (\bibinfo {year} {2009})}\BibitemShut
  {NoStop}%
\bibitem [{\citenamefont {Perdew}\ \emph {et~al.}(1996)\citenamefont {Perdew},
  \citenamefont {Burke},\ and\ \citenamefont
  {Ernzerhof}}]{perdew1996generalized}%
  \BibitemOpen
  \bibfield  {author} {\bibinfo {author} {\bibfnamefont {J.~P.}\ \bibnamefont
  {Perdew}}, \bibinfo {author} {\bibfnamefont {K.}~\bibnamefont {Burke}},\ and\
  \bibinfo {author} {\bibfnamefont {M.}~\bibnamefont {Ernzerhof}},\ }\bibfield
  {title} {\bibinfo {title} {Generalized gradient approximation made simple},\
  }\href@noop {} {\bibfield  {journal} {\bibinfo  {journal} {Phys. Rev. Lett.}\ }\textbf {\bibinfo {volume} {77}},\ \bibinfo {pages} {3865}
  (\bibinfo {year} {1996})}\BibitemShut {NoStop}%
\bibitem [{\citenamefont {Hamann}(2013)}]{hamann2013optimized}%
  \BibitemOpen
  \bibfield  {author} {\bibinfo {author} {\bibfnamefont {D.}~\bibnamefont
  {Hamann}},\ }\bibfield  {title} {\bibinfo {title} {Optimized norm-conserving
  vanderbilt pseudopotentials},\ }\href@noop {} {\bibfield  {journal} {\bibinfo
   {journal} {Phys. Rev. B}\ }\textbf {\bibinfo {volume} {88}},\ \bibinfo
  {pages} {085117} (\bibinfo {year} {2013})}\BibitemShut {NoStop}%
\bibitem [{\citenamefont {Marzari}\ and\ \citenamefont
  {Vanderbilt}(1997)}]{marzari1997maximally}%
  \BibitemOpen
  \bibfield  {author} {\bibinfo {author} {\bibfnamefont {N.}~\bibnamefont
  {Marzari}}\ and\ \bibinfo {author} {\bibfnamefont {D.}~\bibnamefont
  {Vanderbilt}},\ }\bibfield  {title} {\bibinfo {title} {Maximally localized
  generalized \ce{Wannier} functions for composite energy bands},\ }\href@noop
  {} {\bibfield  {journal} {\bibinfo  {journal} {Phys. Rev. B}\ }\textbf
  {\bibinfo {volume} {56}},\ \bibinfo {pages} {12847} (\bibinfo {year}
  {1997})}\BibitemShut {NoStop}%
\bibitem [{\citenamefont {Souza}\ \emph {et~al.}(2001)\citenamefont {Souza},
  \citenamefont {Marzari},\ and\ \citenamefont
  {Vanderbilt}}]{souza2001maximally}%
  \BibitemOpen
  \bibfield  {author} {\bibinfo {author} {\bibfnamefont {I.}~\bibnamefont
  {Souza}}, \bibinfo {author} {\bibfnamefont {N.}~\bibnamefont {Marzari}},\
  and\ \bibinfo {author} {\bibfnamefont {D.}~\bibnamefont {Vanderbilt}},\
  }\bibfield  {title} {\bibinfo {title} {Maximally localized Wannier functions
  for entangled energy bands},\ }\href@noop {} {\bibfield  {journal} {\bibinfo
  {journal} {Phys. Rev. B}\ }\textbf {\bibinfo {volume} {65}},\ \bibinfo
  {pages} {035109} (\bibinfo {year} {2001})}\BibitemShut {NoStop}%
\bibitem [{\citenamefont {Gupta}\ \emph {et~al.}()\citenamefont {Gupta},
  \citenamefont {Husain}, \citenamefont {Kumar}, \citenamefont {Brucas},
  \citenamefont {Rydberg},\ and\ \citenamefont
  {Svedlindh}}]{adom}%
  \BibitemOpen
  \bibfield  {author} {\bibinfo {author} {\bibfnamefont {R.}~\bibnamefont
  {Gupta}}, \bibinfo {author} {\bibfnamefont {S.}~\bibnamefont {Husain}},
  \bibinfo {author} {\bibfnamefont {A.}~\bibnamefont {Kumar}}, \bibinfo
  {author} {\bibfnamefont {R.}~\bibnamefont {Brucas}}, \bibinfo {author}
  {\bibfnamefont {A.}~\bibnamefont {Rydberg}},\ and\ \bibinfo {author}
  {\bibfnamefont {P.}~\bibnamefont {Svedlindh}},\ }\bibfield  {title} {\bibinfo
  {title} {Co$_2$FeAl full \ce{Heusler} compound based spintronic terahertz emitter},\
  }\href {https://doi.org/https://doi.org/10.1002/adom.202001987} {\bibfield
  {journal} {\bibinfo  {journal} {Adv. Opt. Mater.}\ }\textbf
  {\bibinfo {volume} {9}},\ \bibinfo {pages} {2001987}} (\bibinfo {year}
  {2021})\BibitemShut
  {NoStop}%
\bibitem [{\citenamefont {Hazra}\ \emph {et~al.}(2018)\citenamefont {Hazra},
  \citenamefont {Raja}, \citenamefont {Rawat}, \citenamefont {Lakhani},
  \citenamefont {Kaul},\ and\ \citenamefont {Srinath}}]{hazra2018effect}%
  \BibitemOpen
  \bibfield  {author} {\bibinfo {author} {\bibfnamefont {B.~K.}\ \bibnamefont
  {Hazra}}, \bibinfo {author} {\bibfnamefont {M.~M.}\ \bibnamefont {Raja}},
  \bibinfo {author} {\bibfnamefont {R.}~\bibnamefont {Rawat}}, \bibinfo
  {author} {\bibfnamefont {A.}~\bibnamefont {Lakhani}}, \bibinfo {author}
  {\bibfnamefont {S.}~\bibnamefont {Kaul}},\ and\ \bibinfo {author}
  {\bibfnamefont {S.}~\bibnamefont {Srinath}},\ }\bibfield  {title} {\bibinfo
  {title} {Effect of disorder on the anomalous \ce{Hall} conductivity of
  \ce{Co$_2$Fesi} thin films},\ }\href@noop {} {\bibfield  {journal} {\bibinfo
  {journal} {J. Magn. Magn. Mater.}\ }\textbf {\bibinfo
  {volume} {448}},\ \bibinfo {pages} {371} (\bibinfo {year}
  {2018})}\BibitemShut {NoStop}%
\bibitem [{\citenamefont {Kraus}\ and\ \citenamefont
  {Nolze}(1996)}]{kraus1996powder}%
  \BibitemOpen
  \bibfield  {author} {\bibinfo {author} {\bibfnamefont {W.}~\bibnamefont
  {Kraus}}\ and\ \bibinfo {author} {\bibfnamefont {G.}~\bibnamefont {Nolze}},\
  }\bibfield  {title} {\bibinfo {title} {PowderCell-a program for the
  representation and manipulation of crystal structures and calculation of the
  resulting x-ray powder patterns},\ }\href@noop {} {\bibfield  {journal}
  {\bibinfo  {journal} {J. Appl. Crystallogr.}\ }\textbf {\bibinfo
  {volume} {29}},\ \bibinfo {pages} {301} (\bibinfo {year} {1996})}\BibitemShut
  {NoStop}%
\bibitem [{\citenamefont {Ortiz}\ \emph {et~al.}(2011)\citenamefont {Ortiz},
  \citenamefont {Gabor}, \citenamefont {Petrisor}, \citenamefont {Boust},
  \citenamefont {Issac}, \citenamefont {Tiusan}, \citenamefont {Hehn},\ and\
  \citenamefont {Bobo}}]{ortiz2011static}%
  \BibitemOpen
  \bibfield  {author} {\bibinfo {author} {\bibfnamefont {G.}~\bibnamefont
  {Ortiz}}, \bibinfo {author} {\bibfnamefont {M.}~\bibnamefont {Gabor}},
  \bibinfo {author} {\bibfnamefont {T.}~\bibnamefont {Petrisor}, \bibfnamefont
  {Jr}}, \bibinfo {author} {\bibfnamefont {F.}~\bibnamefont {Boust}}, \bibinfo
  {author} {\bibfnamefont {F.}~\bibnamefont {Issac}}, \bibinfo {author}
  {\bibfnamefont {C.}~\bibnamefont {Tiusan}}, \bibinfo {author} {\bibfnamefont
  {M.}~\bibnamefont {Hehn}},\ and\ \bibinfo {author} {\bibfnamefont
  {J.}~\bibnamefont {Bobo}},\ }\bibfield  {title} {\bibinfo {title} {Static and
  dynamic magnetic properties of epitaxial \ce{Co$_2$FeAl} \ce{Heusler} alloy
  thin films},\ }\href@noop {} {\bibfield  {journal} {\bibinfo  {journal}
  {J.  Appl.  Phys.}\ }\textbf {\bibinfo {volume} {109}},\ \bibinfo
  {pages} {07D324} (\bibinfo {year} {2011})}\BibitemShut {NoStop}%
\bibitem [{\citenamefont {Wurmehl}\ \emph {et~al.}(2006)\citenamefont
  {Wurmehl}, \citenamefont {Fecher}, \citenamefont {Kroth}, \citenamefont
  {Kronast}, \citenamefont {D{\"u}rr}, \citenamefont {Takeda}, \citenamefont
  {Saitoh}, \citenamefont {Kobayashi}, \citenamefont {Lin}, \citenamefont
  {Sch{\"o}nhense} \emph {et~al.}}]{wurmehl2006electronic}%
  \BibitemOpen
  \bibfield  {author} {\bibinfo {author} {\bibfnamefont {S.}~\bibnamefont
  {Wurmehl}}, \bibinfo {author} {\bibfnamefont {G.~H.}\ \bibnamefont {Fecher}},
  \bibinfo {author} {\bibfnamefont {K.}~\bibnamefont {Kroth}}, \bibinfo
  {author} {\bibfnamefont {F.}~\bibnamefont {Kronast}}, \bibinfo {author}
  {\bibfnamefont {H.~A.}\ \bibnamefont {D{\"u}rr}}, \bibinfo {author}
  {\bibfnamefont {Y.}~\bibnamefont {Takeda}}, \bibinfo {author} {\bibfnamefont
  {Y.}~\bibnamefont {Saitoh}}, \bibinfo {author} {\bibfnamefont
  {K.}~\bibnamefont {Kobayashi}}, \bibinfo {author} {\bibfnamefont {H.-J.}\
  \bibnamefont {Lin}}, \bibinfo {author} {\bibfnamefont {G.}~\bibnamefont
  {Sch{\"o}nhense}},\ and\ \bibinfo {author} {\bibfnamefont {C.}~\bibnamefont
  {Felser}},\ }\bibfield  {title} {\bibinfo {title}
  {Electronic structure and spectroscopy of the quaternary \ce{Heusler} alloy
  \ce{Co$_2$Cr$_{1- x}$Fe$_x$Al}},\ }\href@noop {} {\bibfield  {journal}
  {\bibinfo  {journal} {J. Phys. D: Appl. Phys.}\ }\textbf
  {\bibinfo {volume} {39}},\ \bibinfo {pages} {803} (\bibinfo {year}
  {2006})}\BibitemShut {NoStop}%
\bibitem [{\citenamefont {Jain}\ \emph {et~al.}(2014)\citenamefont {Jain},
  \citenamefont {Jain}, \citenamefont {Sudheesh}, \citenamefont {Lakshmi},\
  and\ \citenamefont {Venugopalan}}]{jain2014electronic}%
  \BibitemOpen
  \bibfield  {author} {\bibinfo {author} {\bibfnamefont {V.}~\bibnamefont
  {Jain}}, \bibinfo {author} {\bibfnamefont {V.}~\bibnamefont {Jain}}, \bibinfo
  {author} {\bibfnamefont {V.}~\bibnamefont {Sudheesh}}, \bibinfo {author}
  {\bibfnamefont {N.}~\bibnamefont {Lakshmi}},\ and\ \bibinfo {author}
  {\bibfnamefont {K.}~\bibnamefont {Venugopalan}},\ }\bibfield  {title}
  {\bibinfo {title} {Electronic structure and magnetic properties of disordered
  \ce{Co$_2$FeAl} \ce{Heusler} alloy},\ }in\ \href@noop {} {\emph {\bibinfo
  {booktitle} {AIP Conf Proc.}}}{
  \bibinfo {pages} {1544-1545}} (\bibinfo {year} {2014})\BibitemShut {NoStop}%
\bibitem [{\citenamefont {Zhang}\ \emph
  {et~al.}(2018{\natexlab{b}})\citenamefont {Zhang}, \citenamefont {Liu},
  \citenamefont {Yan}, \citenamefont {Niu}, \citenamefont {Lai}, \citenamefont
  {Zhao}, \citenamefont {Wang}, \citenamefont {He}, \citenamefont {Meng},\ and\
  \citenamefont {Xu}}]{zhang2018atomic}%
  \BibitemOpen
  \bibfield  {author} {\bibinfo {author} {\bibfnamefont {X.}~\bibnamefont
  {Zhang}}, \bibinfo {author} {\bibfnamefont {W.}~\bibnamefont {Liu}}, \bibinfo
  {author} {\bibfnamefont {Y.}~\bibnamefont {Yan}}, \bibinfo {author}
  {\bibfnamefont {W.}~\bibnamefont {Niu}}, \bibinfo {author} {\bibfnamefont
  {B.}~\bibnamefont {Lai}}, \bibinfo {author} {\bibfnamefont {Y.}~\bibnamefont
  {Zhao}}, \bibinfo {author} {\bibfnamefont {W.}~\bibnamefont {Wang}}, \bibinfo
  {author} {\bibfnamefont {L.}~\bibnamefont {He}}, \bibinfo {author}
  {\bibfnamefont {H.}~\bibnamefont {Meng}},\ and\ \bibinfo {author}
  {\bibfnamefont {Y.}~\bibnamefont {Xu}},\ }\bibfield  {title} {\bibinfo
  {title} {The atomic-scale magnetism of \ce{Co$_2$FeAl} \ce{Heusler} alloy
  epitaxial thin films},\ }\href@noop {} {\bibfield  {journal} {\bibinfo
  {journal} {Appl. Phys. Lett.}\ }\textbf {\bibinfo {volume} {113}},\
  \bibinfo {pages} {212401} (\bibinfo {year} {2018}{\natexlab{b}})}\BibitemShut
  {NoStop}%
\bibitem [{\citenamefont {Ahmad}\ \emph {et~al.}(2019)\citenamefont {Ahmad},
  \citenamefont {Mitra}, \citenamefont {Srivastava},\ and\ \citenamefont
  {Das}}]{ahmad2019size}%
  \BibitemOpen
  \bibfield  {author} {\bibinfo {author} {\bibfnamefont {A.}~\bibnamefont
  {Ahmad}}, \bibinfo {author} {\bibfnamefont {S.}~\bibnamefont {Mitra}},
  \bibinfo {author} {\bibfnamefont {S.}~\bibnamefont {Srivastava}},\ and\
  \bibinfo {author} {\bibfnamefont {A.}~\bibnamefont {Das}},\ }\bibfield
  {title} {\bibinfo {title} {Size-dependent structural and magnetic properties
  of disordered \ce{Co$_2$FeAl} \ce{Heusler} alloy nanoparticles},\ }\href@noop
  {} {\bibfield  {journal} {\bibinfo  {journal} {J. Magn. Magn. Mater.}\ }\textbf {\bibinfo {volume} {474}},\ \bibinfo {pages}
  {599} (\bibinfo {year} {2019})}\BibitemShut {NoStop}%
\bibitem [{\citenamefont {Hirohata}\ \emph {et~al.}(2006)\citenamefont
  {Hirohata}, \citenamefont {Kikuchi}, \citenamefont {Tezuka}, \citenamefont
  {Inomata}, \citenamefont {Claydon}, \citenamefont {Xu},\ and\ \citenamefont
  {Van~der Laan}}]{hirohata2006heusler}%
  \BibitemOpen
  \bibfield  {author} {\bibinfo {author} {\bibfnamefont {A.}~\bibnamefont
  {Hirohata}}, \bibinfo {author} {\bibfnamefont {M.}~\bibnamefont {Kikuchi}},
  \bibinfo {author} {\bibfnamefont {N.}~\bibnamefont {Tezuka}}, \bibinfo
  {author} {\bibfnamefont {K.}~\bibnamefont {Inomata}}, \bibinfo {author}
  {\bibfnamefont {J.}~\bibnamefont {Claydon}}, \bibinfo {author} {\bibfnamefont
  {Y.}~\bibnamefont {Xu}},\ and\ \bibinfo {author} {\bibfnamefont
  {G.}~\bibnamefont {Van~der Laan}},\ }\bibfield  {title} {\bibinfo {title}
  {\ce{Heusler} alloy/semiconductor hybrid structures},\ }\href@noop {}
  {\bibfield  {journal} {\bibinfo  {journal} {Curr. Opin. Solid State Mater. Sci.}\ }\textbf {\bibinfo {volume} {10}},\ \bibinfo {pages}
  {93} (\bibinfo {year} {2006})}\BibitemShut {NoStop}%
\bibitem [{\citenamefont {Markou}\ \emph {et~al.}(2019)\citenamefont {Markou},
  \citenamefont {Kriegner}, \citenamefont {Gayles}, \citenamefont {Zhang},
  \citenamefont {Chen}, \citenamefont {Ernst}, \citenamefont {Lai},
  \citenamefont {Schnelle}, \citenamefont {Chu}, \citenamefont {Sun} \emph
  {et~al.}}]{markou2019thickness}%
  \BibitemOpen
  \bibfield  {author} {\bibinfo {author} {\bibfnamefont {A.}~\bibnamefont
  {Markou}}, \bibinfo {author} {\bibfnamefont {D.}~\bibnamefont {Kriegner}},
  \bibinfo {author} {\bibfnamefont {J.}~\bibnamefont {Gayles}}, \bibinfo
  {author} {\bibfnamefont {L.}~\bibnamefont {Zhang}}, \bibinfo {author}
  {\bibfnamefont {Y.-C.}\ \bibnamefont {Chen}}, \bibinfo {author}
  {\bibfnamefont {B.}~\bibnamefont {Ernst}}, \bibinfo {author} {\bibfnamefont
  {Y.-H.}\ \bibnamefont {Lai}}, \bibinfo {author} {\bibfnamefont
  {W.}~\bibnamefont {Schnelle}}, \bibinfo {author} {\bibfnamefont {Y.-H.}\
  \bibnamefont {Chu}}, \bibinfo {author} {\bibfnamefont {Y.}~\bibnamefont
  {Sun}},\ and\ \bibinfo {author} {\bibfnamefont {C.}~\bibnamefont
  {Felser}},\ }\bibfield  {title} {\bibinfo {title} {Thickness
  dependence of the anomalous Hall effect in thin films of the topological
  semimetal Co$_2$MnGa},\ }\href@noop {} {\bibfield  {journal} {\bibinfo
  {journal} {Phys. Rev. B}\ }\textbf {\bibinfo {volume} {100}},\ \bibinfo
  {pages} {054422} (\bibinfo {year} {2019})}\BibitemShut {NoStop}%
\bibitem [{\citenamefont {Nozieres}\ and\ \citenamefont
  {Lewiner}(1973)}]{nozieres1973simple}%
  \BibitemOpen
  \bibfield  {author} {\bibinfo {author} {\bibfnamefont {P.}~\bibnamefont
  {Nozieres}}\ and\ \bibinfo {author} {\bibfnamefont {C.}~\bibnamefont
  {Lewiner}},\ }\bibfield  {title} {\bibinfo {title} {A simple theory of the
  anomalous \ce{Hall} effect in semiconductors},\ }\href@noop {} {\bibfield
  {journal} {\bibinfo  {journal} {Journal de Physique}\ }\textbf {\bibinfo
  {volume} {34}},\ \bibinfo {pages} {901} (\bibinfo {year} {1973})}\BibitemShut
  {NoStop}%
\bibitem [{\citenamefont {Onoda}\ \emph {et~al.}(2006)\citenamefont {Onoda},
  \citenamefont {Sugimoto},\ and\ \citenamefont
  {Nagaosa}}]{PhysRevLett.97.126602}%
  \BibitemOpen
  \bibfield  {author} {\bibinfo {author} {\bibfnamefont {S.}~\bibnamefont
  {Onoda}}, \bibinfo {author} {\bibfnamefont {N.}~\bibnamefont {Sugimoto}},\
  and\ \bibinfo {author} {\bibfnamefont {N.}~\bibnamefont {Nagaosa}},\
  }\bibfield  {title} {\bibinfo {title} {Intrinsic versus extrinsic anomalous
  \ce{Hall} effect in ferromagnets},\ }\href
  {https://doi.org/10.1103/PhysRevLett.97.126602} {\bibfield  {journal}
  {\bibinfo  {journal} {Phys. Rev. Lett.}\ }\textbf {\bibinfo {volume} {97}},\
  \bibinfo {pages} {126602} (\bibinfo {year} {2006})}\BibitemShut {NoStop}%
\bibitem [{\citenamefont {Manna}\ \emph
  {et~al.}(2018{\natexlab{b}})\citenamefont {Manna}, \citenamefont {Muechler},
  \citenamefont {Kao}, \citenamefont {Stinshoff}, \citenamefont {Zhang},
  \citenamefont {Gooth}, \citenamefont {Kumar}, \citenamefont {Kreiner},
  \citenamefont {Koepernik}, \citenamefont {Car} \emph
  {et~al.}}]{manna2018colossal}%
  \BibitemOpen
  \bibfield  {author} {\bibinfo {author} {\bibfnamefont {K.}~\bibnamefont
  {Manna}}, \bibinfo {author} {\bibfnamefont {L.}~\bibnamefont {Muechler}},
  \bibinfo {author} {\bibfnamefont {T.-H.}\ \bibnamefont {Kao}}, \bibinfo
  {author} {\bibfnamefont {R.}~\bibnamefont {Stinshoff}}, \bibinfo {author}
  {\bibfnamefont {Y.}~\bibnamefont {Zhang}}, \bibinfo {author} {\bibfnamefont
  {J.}~\bibnamefont {Gooth}}, \bibinfo {author} {\bibfnamefont
  {N.}~\bibnamefont {Kumar}}, \bibinfo {author} {\bibfnamefont
  {G.}~\bibnamefont {Kreiner}}, \bibinfo {author} {\bibfnamefont
  {K.}~\bibnamefont {Koepernik}}, \bibinfo {author} {\bibfnamefont
  {R.}~\bibnamefont {Car}}, \bibinfo {author} {\bibfnamefont
  {J.}~\bibnamefont {K{\"u}bler}}, \bibinfo {author} {\bibfnamefont
  {G.~H.}~\bibnamefont {Fecher}}, \bibinfo {author} {\bibfnamefont
  {C.}~\bibnamefont {Shekhar}}, \bibinfo {author} {\bibfnamefont
  {Y.}~\bibnamefont {Sun}},\ and\  \bibinfo {author} {\bibfnamefont
  {C.}~\bibnamefont {Felser}}, \ }\bibfield  {title} {\bibinfo
  {title} {From colossal to zero: controlling the anomalous \ce{Hall} effect in
  magnetic \ce{Heusler} compounds via \ce{Berry} curvature design},\ }\href@noop {}
  {\bibfield  {journal} {\bibinfo  {journal} {Phys. Rev. X}\ }\textbf
  {\bibinfo {volume} {8}},\ \bibinfo {pages} {041045} (\bibinfo {year}
  {2018}{\natexlab{b}})}\BibitemShut {NoStop}%
\bibitem [{\citenamefont {Wang}\ \emph {et~al.}(2018)\citenamefont {Wang},
  \citenamefont {Xu}, \citenamefont {Lou}, \citenamefont {Liu}, \citenamefont
  {Li}, \citenamefont {Huang}, \citenamefont {Shen}, \citenamefont {Weng},
  \citenamefont {Wang},\ and\ \citenamefont {Lei}}]{wang2018large}%
  \BibitemOpen
  \bibfield  {author} {\bibinfo {author} {\bibfnamefont {Q.}~\bibnamefont
  {Wang}}, \bibinfo {author} {\bibfnamefont {Y.}~\bibnamefont {Xu}}, \bibinfo
  {author} {\bibfnamefont {R.}~\bibnamefont {Lou}}, \bibinfo {author}
  {\bibfnamefont {Z.}~\bibnamefont {Liu}}, \bibinfo {author} {\bibfnamefont
  {M.}~\bibnamefont {Li}}, \bibinfo {author} {\bibfnamefont {Y.}~\bibnamefont
  {Huang}}, \bibinfo {author} {\bibfnamefont {D.}~\bibnamefont {Shen}},
  \bibinfo {author} {\bibfnamefont {H.}~\bibnamefont {Weng}}, \bibinfo {author}
  {\bibfnamefont {S.}~\bibnamefont {Wang}},\ and\ \bibinfo {author}
  {\bibfnamefont {H.}~\bibnamefont {Lei}},\ }\bibfield  {title} {\bibinfo
  {title} {Large intrinsic anomalous \ce{Hall} effect in half-metallic
  ferromagnet \ce{Co$_3$Sn$_2$S$_2$} with magnetic \ce{Weyl} fermions},\
  }\href@noop {} {\bibfield  {journal} {\bibinfo  {journal} {Nat. Commun.}\ }\textbf {\bibinfo {volume} {9}},\ \bibinfo {pages} {3681}
  (\bibinfo {year} {2018})}\BibitemShut {NoStop}%
\bibitem [{\citenamefont {Liu}\ \emph {et~al.}(2019)\citenamefont {Liu},
  \citenamefont {Liang}, \citenamefont {Liu}, \citenamefont {Xu}, \citenamefont
  {Li}, \citenamefont {Chen}, \citenamefont {Pei}, \citenamefont {Shi},
  \citenamefont {Mo}, \citenamefont {Dudin} \emph {et~al.}}]{liu2019magnetic}%
  \BibitemOpen
  \bibfield  {author} {\bibinfo {author} {\bibfnamefont {D.}~\bibnamefont
  {Liu}}, \bibinfo {author} {\bibfnamefont {A.}~\bibnamefont {Liang}}, \bibinfo
  {author} {\bibfnamefont {E.}~\bibnamefont {Liu}}, \bibinfo {author}
  {\bibfnamefont {Q.}~\bibnamefont {Xu}}, \bibinfo {author} {\bibfnamefont
  {Y.}~\bibnamefont {Li}}, \bibinfo {author} {\bibfnamefont {C.}~\bibnamefont
  {Chen}}, \bibinfo {author} {\bibfnamefont {D.}~\bibnamefont {Pei}}, \bibinfo
  {author} {\bibfnamefont {W.}~\bibnamefont {Shi}}, \bibinfo {author}
  {\bibfnamefont {S.}~\bibnamefont {Mo}}, \bibinfo {author} {\bibfnamefont
  {P.}~\bibnamefont {Dudin}}, \bibinfo {author} {\bibfnamefont
  {P.}~\bibnamefont {Dudin}}, \bibinfo {author} {\bibfnamefont
  {T.}~\bibnamefont {Kim}}, \bibinfo {author} {\bibfnamefont
  {C.}~\bibnamefont {Cacho}}, \bibinfo {author} {\bibfnamefont
  {G.}~\bibnamefont {Li}}, \bibinfo {author} {\bibfnamefont
  {Y.}~\bibnamefont {Sun}}, \bibinfo {author} {\bibfnamefont
  {L.~X.}~\bibnamefont {Yang}}, \bibinfo {author} {\bibfnamefont
  {Z.~K.}~\bibnamefont {Liu}}, \bibinfo {author} {\bibfnamefont
  {S.~S.~P.}~\bibnamefont {Parkin}}, \bibinfo {author} {\bibfnamefont
  {C.}~\bibnamefont {Felser}},\ and\ \bibinfo {author} {\bibfnamefont
  {Y.L.}~\bibnamefont {Chen}},\ }\bibfield  {title} {\bibinfo
  {title} {Magnetic \ce{Weyl} semimetal phase in a kagom{\'e} crystal},\
  }\href@noop {} {\bibfield  {journal} {\bibinfo  {journal} {Science}\ }\textbf
  {\bibinfo {volume} {365}},\ \bibinfo {pages} {1282} (\bibinfo {year}
  {2019})}\BibitemShut {NoStop}%
\bibitem [{\citenamefont {Chen}\ \emph {et~al.}(2019)\citenamefont {Chen},
  \citenamefont {Wang}, \citenamefont {Gu}, \citenamefont {Wang}, \citenamefont
  {Zhou}, \citenamefont {An}, \citenamefont {Zhou}, \citenamefont {Zhang},
  \citenamefont {Chen}, \citenamefont {Yuan} \emph
  {et~al.}}]{chen2019pressure}%
  \BibitemOpen
  \bibfield  {author} {\bibinfo {author} {\bibfnamefont {X.}~\bibnamefont
  {Chen}}, \bibinfo {author} {\bibfnamefont {M.}~\bibnamefont {Wang}}, \bibinfo
  {author} {\bibfnamefont {C.}~\bibnamefont {Gu}}, \bibinfo {author}
  {\bibfnamefont {S.}~\bibnamefont {Wang}}, \bibinfo {author} {\bibfnamefont
  {Y.}~\bibnamefont {Zhou}}, \bibinfo {author} {\bibfnamefont {C.}~\bibnamefont
  {An}}, \bibinfo {author} {\bibfnamefont {Y.}~\bibnamefont {Zhou}}, \bibinfo
  {author} {\bibfnamefont {B.}~\bibnamefont {Zhang}}, \bibinfo {author}
  {\bibfnamefont {C.}~\bibnamefont {Chen}}, \bibinfo {author} {\bibfnamefont
  {Y.}~\bibnamefont {Yuan}}, \bibinfo {author} {\bibfnamefont
  {M.}~\bibnamefont {Qi}}, \bibinfo {author} {\bibfnamefont
  {L.}~\bibnamefont {Zhang}}, \bibinfo {author} {\bibfnamefont
  {H.}~\bibnamefont {Zhou}}, \bibinfo {author} {\bibfnamefont
  {J.}~\bibnamefont {Zhou}}, \bibinfo {author} {\bibfnamefont
  {Y.}~\bibnamefont {Yao}},\ and\ \bibinfo {author} {\bibfnamefont
  {Z.}~\bibnamefont {Yang}},\ }\bibfield  {title} {\bibinfo
  {title} {Pressure-tunable large anomalous \ce{Hall} effect of the
  ferromagnetic kagome-lattice \ce{Weyl} semimetal \ce{Co$_3$Sn$_2$S$_2$}},\
  }\href@noop {} {\bibfield  {journal} {\bibinfo  {journal} {Phys. Rev. B}\ }\textbf {\bibinfo {volume} {100}},\ \bibinfo {pages} {165145} (\bibinfo
  {year} {2019})}\BibitemShut {NoStop}%
 \bibitem [{\citenamefont {Chaudhary}\ \emph {et~al.}(2021)\citenamefont
  {Chaudhary}, \citenamefont {Dubey}, \citenamefont {Shukla}, \citenamefont
  {Singh}, \citenamefont {Sadhukhan}, \citenamefont {Kanungo}, \citenamefont
  {Jena}, \citenamefont {Lee}, \citenamefont {Bhattacharjee}, \citenamefont
  {Min\'ar},\ and\ \citenamefont {D'Souza}}]{CTS}%
  \BibitemOpen
  \bibfield  {author} {\bibinfo {author} {\bibfnamefont {P.}~\bibnamefont
  {Chaudhary}}, \bibinfo {author} {\bibfnamefont {K.~K.}\ \bibnamefont
  {Dubey}}, \bibinfo {author} {\bibfnamefont {G.~K.}\ \bibnamefont {Shukla}},
  \bibinfo {author} {\bibfnamefont {S.}~\bibnamefont {Singh}}, \bibinfo
  {author} {\bibfnamefont {S.}~\bibnamefont {Sadhukhan}}, \bibinfo {author}
  {\bibfnamefont {S.}~\bibnamefont {Kanungo}}, \bibinfo {author} {\bibfnamefont
  {A.~K.}\ \bibnamefont {Jena}}, \bibinfo {author} {\bibfnamefont {S.-C.}\
  \bibnamefont {Lee}}, \bibinfo {author} {\bibfnamefont {S.}~\bibnamefont
  {Bhattacharjee}}, \bibinfo {author} {\bibfnamefont {J.}~\bibnamefont
  {Min\'ar}},\ and\ \bibinfo {author} {\bibfnamefont {S.~W.}\ \bibnamefont
  {D'Souza}},\ }\bibfield  {title} {\bibinfo {title} {Role of chemical disorder
  in tuning the Weyl points in vanadium doped Co$_2$TiSn},\ }\href
  {https://doi.org/10.1103/PhysRevMaterials.5.124201} {\bibfield  {journal}
  {\bibinfo  {journal} {Phys. Rev. Mater.}\ }\textbf {\bibinfo {volume}
  {5}},\ \bibinfo {pages} {124201} (\bibinfo {year} {2021})}\BibitemShut
  {NoStop}%
\bibitem [{\citenamefont {Wang}\ \emph {et~al.}(2016)\citenamefont {Wang},
  \citenamefont {Vergniory}, \citenamefont {Kushwaha}, \citenamefont
  {Hirschberger}, \citenamefont {Chulkov}, \citenamefont {Ernst}, \citenamefont
  {Ong}, \citenamefont {Cava},\ and\ \citenamefont {Bernevig}}]{Weyl}%
  \BibitemOpen
  \bibfield  {author} {\bibinfo {author} {\bibfnamefont {Z.}~\bibnamefont
  {Wang}}, \bibinfo {author} {\bibfnamefont {M.~G.}\ \bibnamefont {Vergniory}},
  \bibinfo {author} {\bibfnamefont {S.}~\bibnamefont {Kushwaha}}, \bibinfo
  {author} {\bibfnamefont {M.}~\bibnamefont {Hirschberger}}, \bibinfo {author}
  {\bibfnamefont {E.~V.}\ \bibnamefont {Chulkov}}, \bibinfo {author}
  {\bibfnamefont {A.}~\bibnamefont {Ernst}}, \bibinfo {author} {\bibfnamefont
  {N.~P.}\ \bibnamefont {Ong}}, \bibinfo {author} {\bibfnamefont {R.~J.}\
  \bibnamefont {Cava}},\ and\ \bibinfo {author} {\bibfnamefont {B.~A.}\
  \bibnamefont {Bernevig}},\ }\bibfield  {title} {\bibinfo {title}
  {Time-reversal-breaking Weyl fermions in magnetic Heusler alloys},\ }\href
  {https://doi.org/10.1103/PhysRevLett.117.236401} {\bibfield  {journal}
  {\bibinfo  {journal} {Phys. Rev. Lett.}\ }\textbf {\bibinfo {volume} {117}},\
  \bibinfo {pages} {236401} (\bibinfo {year} {2016})}\BibitemShut {NoStop}%
 \bibitem [{\citenamefont {K{\"u}bler}\ and\ \citenamefont
  {Felser}(2012)}]{kubler2012berry}%
  \BibitemOpen
  \bibfield  {author} {\bibinfo {author} {\bibfnamefont {J.}~\bibnamefont
  {K{\"u}bler}}\ and\ \bibinfo {author} {\bibfnamefont {C.}~\bibnamefont
  {Felser}},\ }\bibfield  {title} {\bibinfo {title} {\ce{Berry} curvature and
  the anomalous \ce{Hall} effect in \ce{Heusler} compounds},\ }\href@noop {}
  {\bibfield  {journal} {\bibinfo  {journal} {Physical Review B}\ }\textbf
  {\bibinfo {volume} {85}},\ \bibinfo {pages} {012405} (\bibinfo {year}
  {2012})}\BibitemShut {NoStop}%
\bibitem [{\citenamefont {Pizzi}\ \emph {et~al.}(2020)\citenamefont {Pizzi},
  \citenamefont {Vitale}, \citenamefont {Arita}, \citenamefont {Blügel},
  \citenamefont {Freimuth}, \citenamefont {G{\'{e}}ranton}, \citenamefont
  {Gibertini}, \citenamefont {Gresch}, \citenamefont {Johnson}, \citenamefont
  {Koretsune}, \citenamefont {Iba{\~{n}}ez-Azpiroz}, \citenamefont {Lee},
  \citenamefont {Lihm}, \citenamefont {Marchand}, \citenamefont {Marrazzo},
  \citenamefont {Mokrousov}, \citenamefont {Mustafa}, \citenamefont {Nohara},
  \citenamefont {Nomura}, \citenamefont {Paulatto}, \citenamefont
  {Ponc{\'{e}}}, \citenamefont {Ponweiser}, \citenamefont {Qiao}, \citenamefont
  {Thöle}, \citenamefont {Tsirkin}, \citenamefont {Wierzbowska}, \citenamefont
  {Marzari}, \citenamefont {Vanderbilt}, \citenamefont {Souza}, \citenamefont
  {Mostofi},\ and\ \citenamefont {Yates}}]{pizzi2020wannier90}%
  \BibitemOpen
  \bibfield  {author} {\bibinfo {author} {\bibfnamefont {G.}~\bibnamefont
  {Pizzi}}, \bibinfo {author} {\bibfnamefont {V.}~\bibnamefont {Vitale}},
  \bibinfo {author} {\bibfnamefont {R.}~\bibnamefont {Arita}}, \bibinfo
  {author} {\bibfnamefont {S.}~\bibnamefont {Blügel}}, \bibinfo {author}
  {\bibfnamefont {F.}~\bibnamefont {Freimuth}}, \bibinfo {author}
  {\bibfnamefont {G.}~\bibnamefont {G{\'{e}}ranton}}, \bibinfo {author}
  {\bibfnamefont {M.}~\bibnamefont {Gibertini}}, \bibinfo {author}
  {\bibfnamefont {D.}~\bibnamefont {Gresch}}, \bibinfo {author} {\bibfnamefont
  {C.}~\bibnamefont {Johnson}}, \bibinfo {author} {\bibfnamefont
  {T.}~\bibnamefont {Koretsune}}, \bibinfo {author} {\bibfnamefont
  {J.}~\bibnamefont {Iba{\~{n}}ez-Azpiroz}}, \bibinfo {author} {\bibfnamefont
  {H.}~\bibnamefont {Lee}}, \bibinfo {author} {\bibfnamefont {J.-M.}\
  \bibnamefont {Lihm}}, \bibinfo {author} {\bibfnamefont {D.}~\bibnamefont
  {Marchand}}, \bibinfo {author} {\bibfnamefont {A.}~\bibnamefont {Marrazzo}},
  \bibinfo {author} {\bibfnamefont {Y.}~\bibnamefont {Mokrousov}}, \bibinfo
  {author} {\bibfnamefont {J.~I.}\ \bibnamefont {Mustafa}}, \bibinfo {author}
  {\bibfnamefont {Y.}~\bibnamefont {Nohara}}, \bibinfo {author} {\bibfnamefont
  {Y.}~\bibnamefont {Nomura}}, \bibinfo {author} {\bibfnamefont
  {L.}~\bibnamefont {Paulatto}}, \bibinfo {author} {\bibfnamefont
  {S.}~\bibnamefont {Ponc{\'{e}}}}, \bibinfo {author} {\bibfnamefont
  {T.}~\bibnamefont {Ponweiser}}, \bibinfo {author} {\bibfnamefont
  {J.}~\bibnamefont {Qiao}}, \bibinfo {author} {\bibfnamefont {F.}~\bibnamefont
  {Thöle}}, \bibinfo {author} {\bibfnamefont {S.~S.}\ \bibnamefont {Tsirkin}},
  \bibinfo {author} {\bibfnamefont {M.}~\bibnamefont {Wierzbowska}}, \bibinfo
  {author} {\bibfnamefont {N.}~\bibnamefont {Marzari}}, \bibinfo {author}
  {\bibfnamefont {D.}~\bibnamefont {Vanderbilt}}, \bibinfo {author}
  {\bibfnamefont {I.}~\bibnamefont {Souza}}, \bibinfo {author} {\bibfnamefont
  {A.~A.}\ \bibnamefont {Mostofi}},\ and\ \bibinfo {author} {\bibfnamefont
  {J.~R.}\ \bibnamefont {Yates}},\ }\bibfield  {title} {\bibinfo {title}
  {Wannier90 as a community code: new features and applications},\ }\href
  {https://doi.org/10.1088/1361-648x/ab51ff} {\bibfield  {journal} {\bibinfo
  {journal} {J. Phys. Condens. Matter}\ }\textbf {\bibinfo {volume} {32}},\
  \bibinfo {pages} {165902} (\bibinfo {year} {2020})}\BibitemShut {NoStop}%
\bibitem [{\citenamefont {Tsirkin}(2021)}]{tsirkin2021high}%
  \BibitemOpen
  \bibfield  {author} {\bibinfo {author} {\bibfnamefont {S.~S.}\ \bibnamefont
  {Tsirkin}},\ }\bibfield  {title} {\bibinfo {title} {High performance Wannier
  interpolation of \ce{Berry} curvature and related quantities with
  WannierBerri code},\ }\href@noop {} {\bibfield  {journal} {\bibinfo
  {journal} {npj Comput. Mater.}\ }\textbf {\bibinfo {volume} {7}},\ \bibinfo
  {pages} {33} (\bibinfo {year} {2021})}\BibitemShut {NoStop}%
\bibitem [{\citenamefont {Mostofi}\ \emph {et~al.}(2008)\citenamefont
  {Mostofi}, \citenamefont {Yates}, \citenamefont {Lee}, \citenamefont {Souza},
  \citenamefont {Vanderbilt},\ and\ \citenamefont {Marzari}}]{MOSTOFI2008685}%
  \BibitemOpen
  \bibfield  {author} {\bibinfo {author} {\bibfnamefont {A.~A.}\ \bibnamefont
  {Mostofi}}, \bibinfo {author} {\bibfnamefont {J.~R.}\ \bibnamefont {Yates}},
  \bibinfo {author} {\bibfnamefont {Y.-S.}\ \bibnamefont {Lee}}, \bibinfo
  {author} {\bibfnamefont {I.}~\bibnamefont {Souza}}, \bibinfo {author}
  {\bibfnamefont {D.}~\bibnamefont {Vanderbilt}},\ and\ \bibinfo {author}
  {\bibfnamefont {N.}~\bibnamefont {Marzari}},\ }\bibfield  {title} {\bibinfo
  {title} {Wannier90: A tool for obtaining maximally-localised \ce{Wannier
  functions}},\ }\href@noop {} {\bibfield  {journal} {\bibinfo  {journal}
  {Comput. Phys. Commun.}\ }\textbf {\bibinfo {volume} {178}},\ \bibinfo
  {pages} {685} (\bibinfo {year} {2008})}\BibitemShut {NoStop}%
\bibitem [{\citenamefont {Marzari}\ and\ \citenamefont
  {Vanderbilt}(1997)}]{PhysRevB.56.12847}%
  \BibitemOpen
  \bibfield  {author} {\bibinfo {author} {\bibfnamefont {N.}~\bibnamefont
  {Marzari}}\ and\ \bibinfo {author} {\bibfnamefont {D.}~\bibnamefont
  {Vanderbilt}},\ }\bibfield  {title} {\bibinfo {title} {Maximally localized
  generalized Wannier functions for composite energy bands},\ }\href
  {https://doi.org/10.1103/PhysRevB.56.12847} {\bibfield  {journal} {\bibinfo
  {journal} {Phys. Rev. B}\ }\textbf {\bibinfo {volume} {56}},\ \bibinfo
  {pages} {12847} (\bibinfo {year} {1997})}\BibitemShut {NoStop}%
 \bibitem [{\citenamefont {Kundu}\ \emph {et~al.}(2021)\citenamefont {Kundu},
  \citenamefont {Bhattacharjee}, \citenamefont {Lee},\ and\ \citenamefont
  {Jain}}]{wannierimp}%
  \BibitemOpen
  \bibfield  {author} {\bibinfo {author} {\bibfnamefont {S.}~\bibnamefont
  {Kundu}}, \bibinfo {author} {\bibfnamefont {S.}~\bibnamefont
  {Bhattacharjee}}, \bibinfo {author} {\bibfnamefont {S.-C.}\ \bibnamefont
  {Lee}},\ and\ \bibinfo {author} {\bibfnamefont {M.}~\bibnamefont {Jain}},\
  }\bibfield  {title} {\bibinfo {title} {Population analysis with Wannier
  orbitals},\ }\href@noop {} {\bibfield  {journal} {\bibinfo  {journal} {J.
  Chem. Phys.}\ }\textbf {\bibinfo {volume} {154}},\ \bibinfo {pages} {104111}
  (\bibinfo {year} {2021})}\BibitemShut {NoStop}%
\bibitem [{\citenamefont {Huang}\ \emph {et~al.}(2015)\citenamefont {Huang},
  \citenamefont {Tung},\ and\ \citenamefont {Guo}}]{huang2015anomalous}%
  \BibitemOpen
  \bibfield  {author} {\bibinfo {author} {\bibfnamefont {H.-L.}\ \bibnamefont
  {Huang}}, \bibinfo {author} {\bibfnamefont {J.-C.}\ \bibnamefont {Tung}},\
  and\ \bibinfo {author} {\bibfnamefont {G.-Y.}\ \bibnamefont {Guo}},\
  }\bibfield  {title} {\bibinfo {title} {Anomalous \ce{Hall} effect and current
  spin polarization in \ce{Co$_2$FeX} \ce{Heusler} compounds \ce{(X= Al, Ga,
  In, Si, Ge, and Sn)}: A systematic ab initio study},\ }\href@noop {}
  {\bibfield  {journal} {\bibinfo  {journal} {Phys. Rev. B}\ }\textbf
  {\bibinfo {volume} {91}},\ \bibinfo {pages} {134409} (\bibinfo {year}
  {2015})}\BibitemShut {NoStop}%
\bibitem [{\citenamefont {Zhu}\ \emph {et~al.}(2020)\citenamefont {Zhu},
  \citenamefont {Singh}, \citenamefont {Wang}, \citenamefont {Huang},
  \citenamefont {Chiu}, \citenamefont {Wang}, \citenamefont {Graf},
  \citenamefont {Zhang}, \citenamefont {Lin}, \citenamefont {Sun} \emph
  {et~al.}}]{zhu2020exceptionally}%
  \BibitemOpen
  \bibfield  {author} {\bibinfo {author} {\bibfnamefont {Y.}~\bibnamefont
  {Zhu}}, \bibinfo {author} {\bibfnamefont {B.}~\bibnamefont {Singh}}, \bibinfo
  {author} {\bibfnamefont {Y.}~\bibnamefont {Wang}}, \bibinfo {author}
  {\bibfnamefont {C.-Y.}\ \bibnamefont {Huang}}, \bibinfo {author}
  {\bibfnamefont {W.-C.}\ \bibnamefont {Chiu}}, \bibinfo {author}
  {\bibfnamefont {B.}~\bibnamefont {Wang}}, \bibinfo {author} {\bibfnamefont
  {D.}~\bibnamefont {Graf}}, \bibinfo {author} {\bibfnamefont {Y.}~\bibnamefont
  {Zhang}}, \bibinfo {author} {\bibfnamefont {H.}~\bibnamefont {Lin}}, \bibinfo
  {author} {\bibfnamefont {J.}~\bibnamefont {Sun}}, 
  \bibinfo
  {author} {\bibfnamefont {A.}~\bibnamefont {Bansil}},\ and\ \bibinfo
  {author} {\bibfnamefont {Z.}~\bibnamefont {Mao}},\ }\bibfield
   {title} {\bibinfo {title} {Exceptionally large anomalous \ce{Hall} effect due to
  anticrossing of spin-split bands in the antiferromagnetic \ce{half-Heusler}
  compound \ce{TbPtBi}},\ }\href@noop {} {\bibfield  {journal} {\bibinfo  {journal}
  {Phys. Rev. B}\ }\textbf {\bibinfo {volume} {101}},\ \bibinfo {pages}
  {161105} (\bibinfo {year} {2020})}\BibitemShut {NoStop}%

\end{thebibliography}
\end{document}